\documentclass[aip,amsmath,amssymb,amsfonts,reprint]{revtex4-1}

\usepackage{graphicx} 
\usepackage{MnSymbol}

\usepackage[draft]{changes}

\usepackage[english]{babel}
\usepackage{ucs}
\usepackage[utf8x]{inputenc}
\usepackage{nicematrix}

\newcommand{\0}{\Bar{0}}
\newcommand{\1}{\Bar{1}}

\newcommand{\N}{\mathbb{N}}

\newtheorem{theor}{Theorem}
\newtheorem{defi}{Definition}

\begin{document}

\title{Templex: a bridge between homologies and templates for chaotic attractors}

\author{Gisela D. Char\'o}
\email{gisela.charo@cima.fcen.uba.ar}
\affiliation{CONICET – Universidad de Buenos Aires. Centro de Investigaciones
del Mar y la Atmósfera (CIMA), C1428EGA  CABA, Argentina}
\affiliation{CNRS – IRD – CONICET – UBA. Institut Franco-Argentin d'\'Etudes sur le Climat et ses Impacts (IRL 3351 IFAECI), C1428EGA  CABA, Argentina.}%

\author{Christophe Letellier}
\homepage{http://www.atomosyd.net/spip.php?article1}
\email{christophe.letellier@coria.fr}

\affiliation{
Rouen Normandie University --- CORIA, Campus Universitaire du Madrillet,
F-76800 Saint-Etienne du Rouvray, France }

%


\author{Denisse Sciamarella}
\email{denisse.sciamarella@cima.fcen.uba.ar}
\affiliation{CNRS – Centre National de la Recherche Scientifique, 75016 Paris, France.}%
\affiliation{CNRS – IRD – CONICET – UBA. Institut Franco-Argentin d'\'Etudes sur le Climat et ses Impacts (IRL 3351 IFAECI), C1428EGA  CABA, Argentina.}

\date{\today ~to submit to {\it Chaos}}

\begin{abstract}

The theory of homologies introduces cell complexes to provide an algebraic description of spaces up to topological equivalence. Attractors in state space can be studied using Branched Manifold Analysis through Homologies: this strategy constructs a cell complex from a cloud of points in state space and uses homology groups to characterize its topology. The approach, however, does not consider the action of the flow on the cell complex. The procedure is here extended to take this fundamental property into account, as done with templates. The goal is achieved endowing the cell complex with a directed graph that prescribes the flow direction between its highest dimensional cells. The tandem of cell complex and directed graph, baptized templex, is shown to allow for a sophisticated characterization of chaotic attractors and for an accurate classification of them. The cases of a few well-known chaotic attractors are investigated --- namely the spiral and funnel R\"ossler attractors, the Lorenz attractor and the Burke and Shaw attractor. A link is established with their description in terms of templates. 
\end{abstract}

\maketitle

\begin{quotation}
Cell complexes can be traced back to Poincaré's papers of 1900, and the study 
of chaotic attractors using cell complexes to the nineties. Since then, 
algebraic topology is seen as the mathematical formalism holding promise for a 
description of chaos beyond three dimensions -- there where templates, 
developed in the eighties to extract the knot content of attractors, cannot go. 
The advent of computational methods with a firm ground in topology has 
given a new thrust to this initiative, which is often applied blindly, without 
a sound comprehension of the information contained in a cell complex. In this work, cell complexes are shown to be enhanced as descriptors of the topology of chaotic attractors, when endowed with a directed graph that carries the information of the flow direction in terms of allowed or forbidden cell connections in a complex. This two companion objects are combined in one which is termed ``templex'', a word resulting from the contraction between template and complex. Indeed, the description offered by a complex is as complete and 
precise as the one obtained with a template, without the burden of dimensionality restrictions.  
\end{quotation}

\section{Introduction}

The fact that some dynamical systems may present complex solutions whose description resists analysis was recognized long ago with the three-body  problem.\cite{Bar97} Once Henri Poincaré understood the inherent complexity combined with a sensitivity to initial conditions of its solutions,\cite{Poi90} he developed the {\it Analysis Situs},\cite{Poi95} while knot theory was maturating.\cite{Tai77,Tai84a,Ale15,Ale26b,Rei27} In his investigation of the solutions to the three-body problem, Poincaré was already thinking in terms of manifolds, leading to the concept of homoclinic orbit and the inextricable entanglement which was finally sketched by Melnikov.\cite{Mel63} Using a representation based on isopleths, Lorenz provided the first interpretation of a solution to a dynamical system in terms of what is now called a branched manifold.\cite{Lor63} Slightly later, the concept of branched manifold was formally introduced by Williams\cite{Wil74} who later made a link with knot theory.\cite{Bir83a} Templates were then used to describe the solution to dynamical systems.\cite{Min92,Ghr97b,Gil98} Templates are viewed as a knot-holder.\cite{Bir83a,Tuf92} Indeed, due to the richness of their structure, chaotic attractors (or ``strange'' in the sense of Ruelle and Takens,\cite{Rue71} that is, not regular or not described by a simple manifold) require a sophisticated approach to capture their specificities and to classify them accurately.\cite{Gil03,Let06a,Let21d}

Homological algebra started in the 19th century, with the work of Riemann 
(1857) and Betti (1871).\cite{Bet71,Poi00,Vie27,Hop28a,Hop28b,May29,Die89,Poi95,Wei97} In 1895, Poincaré introduced the notion of homology numbers and in 
1925, Emmy Noether shifted the attention to the “homology groups” of a 
space.\cite{Noe26}
The basic building blocks in theory of homologies are called cells and are assembled into complexes. A cell complex is hence a sort of layered structure, built up of cells of various dimensions.\cite{Kin93} Such a complex can be constructed from a cloud of points in an arbitrary number of dimensions: the cloud is replaced with a set of glued patches, from which a cell complex \cite{Mun18} can be built. The computation of Betti numbers using this approach was applied to regular attractors, that is, to quasi-periodic regimes.\cite{Mul93} A few years later, the approach was extended to handle clouds of points obtained from time-delay embeddings of experimental time series \cite{Sci99} as well as clouds of points obtained by integrating nonlinear dynamical systems related to simple branched manifolds.\cite{Sci01} In these two works, the cell complex can be non-simplicial (facilitating its  construction) and the description is enriched, going beyond 
Betti numbers. The extended algorithm identifies the $k$-generators of the 
homology groups and introduces orientability chains.\cite{Sci99,Sci01} More 
recent applications of this approach, now called BraMAH (Branched Manifold 
Analysis through Homologies), incorporate the extraction of {\it weak 
boundaries},\cite{Cha20,Cha21} enabling a more precise description of 
four-dimensional manifolds. 

Templates can only be constructed for two-dimensional branched manifolds, while homology groups are defined without dimensional restrictions. The description in terms of templates limits the topological analysis of chaotic attractors to those whose embedding dimension is 3. Since there are many higher-dimensional attractors, it is of primary interest to extend the topological analysis to higher-dimensional attractors. An attempt to extend templates beyond three dimensions was proposed but left without achievements.\cite{Lef13} There is no doubt that the most promising approach is based on homology groups, and before attacking high-dimensional attractors, it is necessary to develop an approach based on complexes and homology groups which provides a description of chaotic attractors at 
the accuracy offered by templates. This is what will be developed in this paper
with two key steps: orienting the cells of the complex taking into account the flow direction and associating a directed graph or digraph to it. 
To the best of our 
knowledge, no previous approach has ever considered a cell complex endowed with 
a directed graph, carrying the information of the flow direction in terms of 
allowed cell connections. To accurately consider the rich structure of 
chaotic attractors in terms of cell complexes, it therefore appeared as a 
requirement to introduce a mathematical object that will hereafter be termed 
``templex''. We will thus extend the numerical procedure initially based on a 
sole complex, to a procedure based on a complex endowed with a digraph, allowing to derive subtemplexes (parts of the original templex) which will play the role of strips in templates.

The aim of this paper is to show how a templex bridges the gap between the 
descriptions of chaotic attractors by homologies and templates. A brief introduction to templates and homology groups is provided in Section \ref{homo} with an introduction to the specific concepts required in our sophisticated strategy. In Section \ref{bramako}, three well-known attractors produced by strongly dissipative systems, namely the R\"ossler, the Lorenz and the Burke and Shaw systems, are extensively treated. Section \ref{conc} gives a conclusion. 

\section{Topological background}
\label{homo}

\subsection{Templates as knot-holders for chaotic attractors}
\label{temhold}

\begin{figure}[ht]
  \centering
  \begin{tabular}{ccc}
    \includegraphics[width=0.08\textwidth]{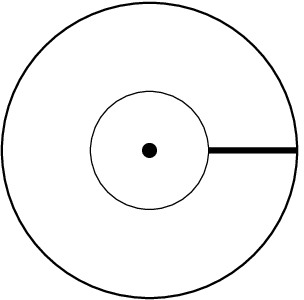} &
    \includegraphics[width=0.18\textwidth]{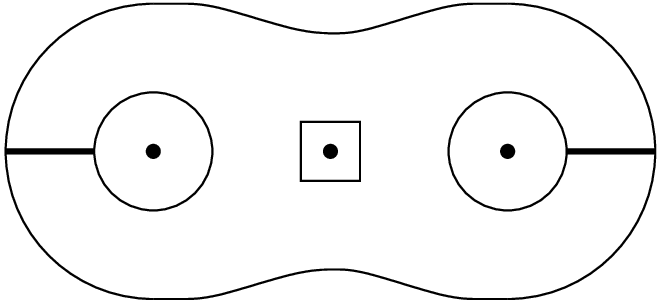} &
    \includegraphics[width=0.18\textwidth]{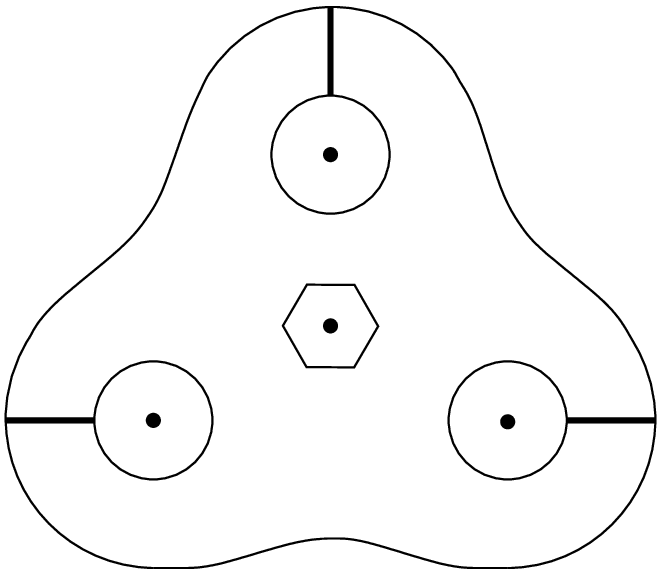} \\
          {\small (a) $g = 1$} & 
	  {\small (b) $g = 3$} &
	  {\small (c) $g = 4$} \\[-0.2cm]
  \end{tabular}
  \caption{Bounding tori of various genus $g$ ($g \leqslant 4$). The $g-1$ components of the Poincaré section are plotted as thick lines. Case (a) applies for the R\"ossler attractor and (b) for the Lorenz attractor. Cases (c) may correspond to the 3-fold covers of the proto-Lorenz system.\cite{Mir93,Let95d} }
  \label{boutos}
\end{figure}

A chaotic attractor is an invariant set under the action of the flow $\phi_t$ which can be bounded by a semi-permeable surface.\cite{Neu00} Consequently, chaotic attractors can be bounded by genus-$g$ tori whose holes are most often associated with singular points circled by the flow.\cite{Tsa03,Tsa04} As we shall see, these holes are not holes in the sense of homologies, that is, they are not necessarily equivalent to generators of the homology groups. They are of two types: (i) those of the focus type which are circled by the flow as a periodic orbit circles a focus point and (ii) those which are associated with a tearing of the flow, splitting the attractor in strips with boundaries. It can be shown that there is always a hole of the saddle type between two holes of the focus type.\cite{Tsa04} A few simple bounding tori are drawn in Fig.\ref{boutos}. The periphery of the attractor is easily defined from the bounding torus and it can be naturally oriented according to the flow. Once the bounding torus is identified, the next step is to construct a Poincaré section and to compute a first-return map to it.  The non trivial result from these bounding tori is that, when $g > 2$, a Poincaré section is made of $g-1$ components\cite{Tsa04} which must be oriented from the center to the periphery to remove some degeneracy among the first-return maps.\cite{Ros13} A bounding torus is thus a manifold which can be naturally oriented according to the flow. 

\begin{figure}[ht]
  \centering
  \begin{tabular}{ccc}
    \includegraphics[width=0.20\textwidth]{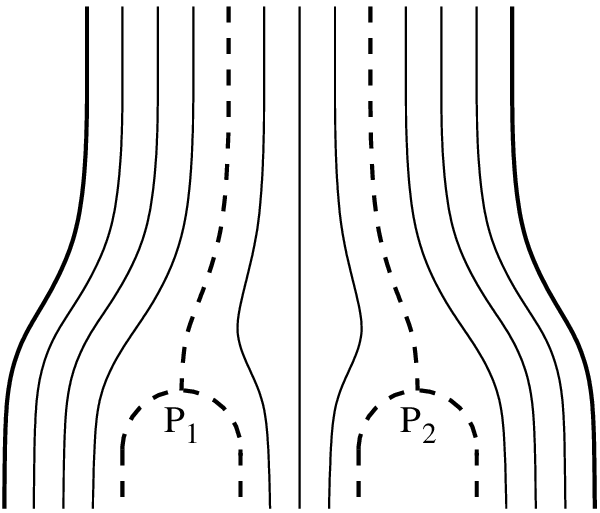} &
    \includegraphics[width=0.18\textwidth]{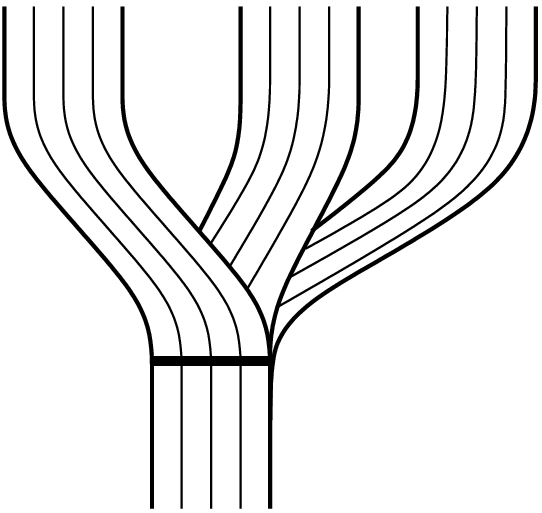} \\
          {\small (a) Splitting chart} &
          {\small (b) Joining chart} \\[-0.2cm]
  \end{tabular}
  \caption{An ingoing strip is split into three outgoing strips according to the two critical points P$_1$ and P$_2$ (a) which are then joined into a single outgoing strip (b). In (a), true boundaries are in solid thick lines and fictive 
boundaries are in dashed thick lines. The very thick line (b) represents the 
joining line (branch line).}
  \label{charts}
\end{figure}

For strongly dissipative systems, the first-return map is one-dimensional and 
the number of monotone branches provides the number $N_{\rm s}$ of strips 
required to construct the corresponding template.\cite{Let95a} The critical 
points of the map --- defining the partition of the map into $N_{\rm s}$ branches --- discriminate the different paths followed by the flow $\phi_t$ determining the (fictive) boundaries between the different strips. Typically, a strip is defined between a splitting chart [Fig.\ \ref{charts}(a)] and a joining chart [Fig.\ \ref{charts}(b)] where the strips are fictively split to allow an easily readable representation of them and where the strips are joined (squeezed) into a single strip, respectively.  According to the standard insertion convention introduced by Tufillaro,\cite{Tuf92} the strips are merged from the back to the front and from the left to the right [Fig.\ \ref{bramaros}(b)]. Typically, all the non-trivial dynamical processes are captured between the splitting chart where the template is split into 
strips and the joining chart where the strips are merged into a single one. The 
joining chart is ended by a joining line which corresponds to a Poincaré 
section [thick line in Fig.\ \ref{charts}(b)]. 

\begin{figure}[ht]
  \centering
  \includegraphics[width=0.45\textwidth]{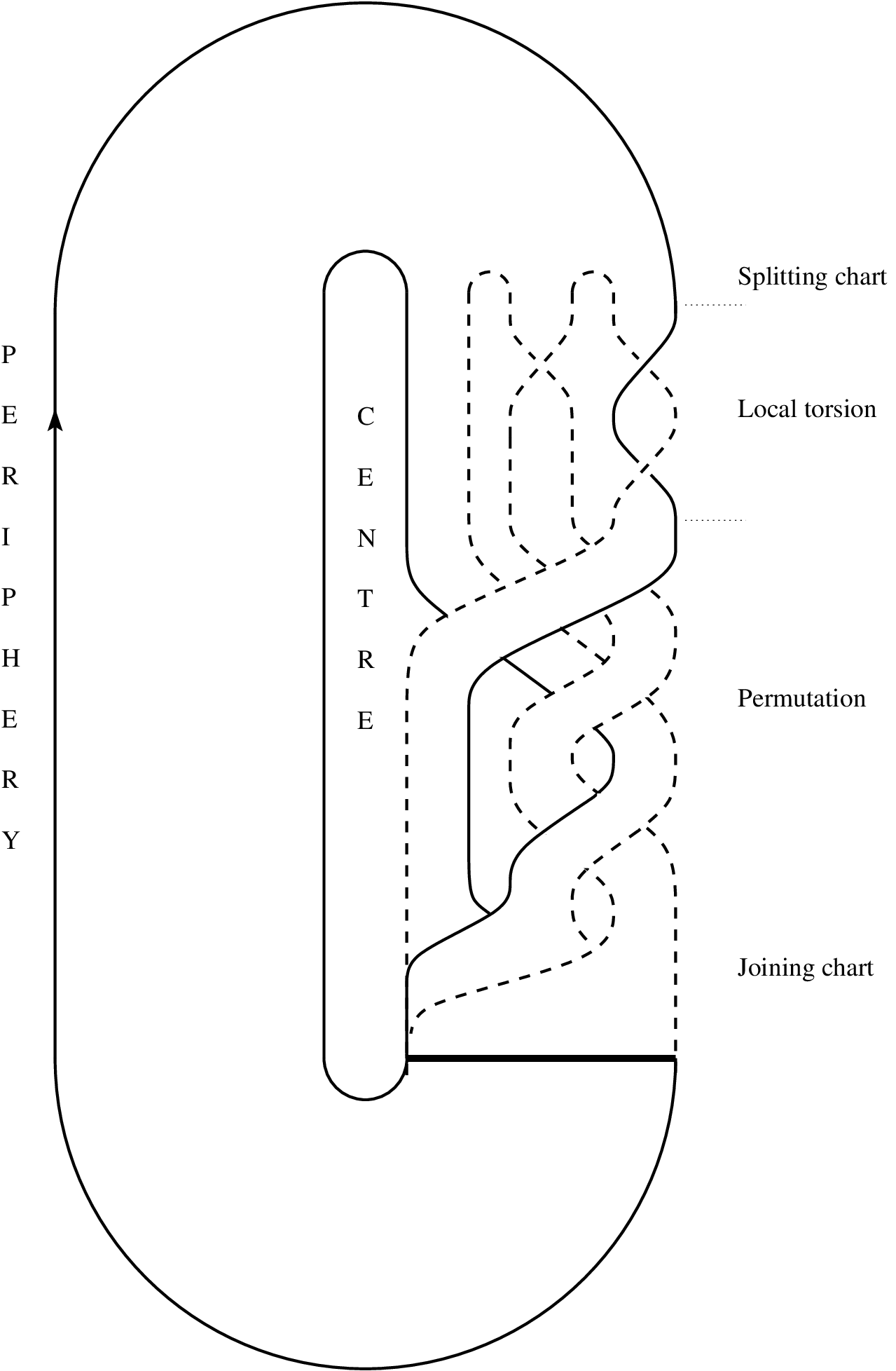} \\[-0.3cm]
  \caption{An example of template made of $N_{\rm s} = 3$ strips and $N_{\rm loc} = 1$ joining line. This template corresponds to an attractor bounded by a 
genus-1 torus. True boundaries are in  solid line and fictive boundaries are in dashed line. The thick line represents  the joining line. The arrow indicates 
the direction of the flow $\phi_t$.}
  \label{extemp}
\end{figure}

Between these two charts, the strips can present local torsion and can be permuted. The template is closed by connecting the joining chart with the splitting chart with a trivial strip as drawn in the example of Fig.\ \ref{extemp}. At the joining chart, there is a joining line\cite{Ghr97b}  (corresponding to the 
joining locus in the next section) 
at which, by definition, the flow cannot be reversed without violating uniqueness. The dynamics is described by the template in terms of charts and strips. According to a theorem due to Birman and Williams, the link of periodic orbits ${\cal L}_\phi$ of the flow $\phi_t$ is in bijective correspondence (under an ambient isotopy) with the link of periodic orbits ${\cal L}_{\rm T}$ of the corresponding template T.\cite{Bir83b}

Each strip can be labelled with an integer using the natural order from the center to the periphery and whose parity is defined by the parity of the local torsion: in the case of Fig.\ \ref{extemp}, we have thus $0 < 1 < 2$ (note that one could have used also $2 < 5 < 8$). When the Poincar\'e section has $N_{\rm c} 
= g -1$ components (joining lines) as for an attractor bounded by a torus with 
a genus $g > 2$, it is still possible to define a {\it total order} 
$\triangleleft$ for ordering the different strips spread in them.\cite{Ghr97b} 
This total order is used to compute the first-return map as it will be 
explained in the case of the Lorenz attractor (see Section \ref{loratt}).

A template can be described using an $N_{\rm s} \times N_{\rm s}$ linking matrix $L_{ij}$ such that $L_{ii}$ is the local torsion of the $i$th strip and $L_{ij}$ is the permutation between the $i$th and the $j$th strips ($i \neq j$).\cite{Tuf92,Let95a} The linking matrix is thus symmetric. When there is a single component to the Poincaré section ($N_{\rm c} = 1$), an $N_{\rm c} \times N_{\rm s}$ joining matrix states the order with which the strips are joined, from the bottom to the top.\cite{Gil98} In the case of the template T drawn in Fig.\ \ref{extemp}, the linking matrix is
\begin{equation}
  L_{ij} = 
  \left[
    \begin{array}{ccc}
       0 & -1 & -1 \\[0.1cm]
      -1 & -1 & -2 \\[0.1cm]
      -1 & -2 & -2 
    \end{array}
  \right\rsem
\end{equation}
and the joining matrix is
\begin{equation}
	J_{ij} = 
  \left[
    \begin{array}{ccc}
       1 & 1 & 1 \\[0.1cm]
    \end{array}
  \right] \, , 
\end{equation}
which is, in this example, trivial and is commonly omitted. Since the links of periodic orbits ${\cal L}_\phi$ and ${\cal L}_{\rm T}$ are equivalent under an ambient isotopy, the linking numbers are equal for both. Linking numbers are defined as\cite{Rol76}
\begin{equation}
  L_{\rm k} = ({\cal O}_1, {\cal O}_2) = \frac{1}{2} \, 
	\sum_{{\cal O}_1 \cap {\cal O}_2} \, \epsilon_i 
\end{equation}
where $\epsilon_i = \pm 1$ is the sign of the $i$th crossing and ${\cal O}_1 \cap {\cal O}_2$ denotes the crossings between an orbit ${\cal O}_1$ and an orbit ${\cal O}_2$ in some regular representation. By definition, $L_{\rm k} ({\cal O}_1, {\cal O}_2) \in \mathbb{Z}$.


\subsection{Homology groups}
\label{homogro}

Topological data analysis through homologies starts with the construction of a complex from a finite set of points in some multidimensional space. In nonlinear dynamics, this set of points can proceed, for instance, from the integration of a system of ordinary differential equations, or from an embedding of a dataset. There are
different rules for constructing a complex. For instance, a Čech complex is a 
complex built on a set of points such that, for balls of a certain radius 
defined around all the points in the set, there is a cell for every finite 
subset of balls with nonempty intersection.\cite{Ghr14} In the case of BraMAH, 
the rules to construct the cell complex consider the set of points in a 
multidimensional state space which lies on a branched $\kappa$-manifold 
($\kappa \in \N$), and uses subsets of points that can be locally approximated 
by $\kappa$-disks to construct $\kappa$-cells [whose boundaries have as many 
$(\kappa -1)$-cells as necessary], and glues them together in a complex of 
dimension $\kappa$. A BraMAH complex is a cell complex built in this manner. By construction, the dimension of a BraMAH complex coincides with the local dimension of the manifold on which an attractor lies. For strongly dissipative dynamical systems as considered in this work, the data points are recorded from the invariant set and they lie on a branched $2$-manifold (an algorithmic procedure to construct a BraMAH complex of dimension $2$ from a cloud of points in three or four dimensions is proposed in Refs. \onlinecite{Sci99,Sci01,Cha20,Cha21,Cha21b}). The present paper focusses, 
not on this construction, but on the extraction of the topological properties 
from a BraMAH complex.

Before defining homology groups, let us first introduce the class of spaces for which they are defined, which is the class of all polyhedra.\cite{Mun18} A polyhedron is a space that can be built from ``building blocks'' as line segments, polygons, polyhedra, and their higher dimensional analogues, by ``gluing them together'' along their faces. A $k$-cell is defined as a set whose interior is homeomorphic to a $k$-dimensional disc with the additional property that its boundary must be divided into a finite number of ($k-1$)-cells, called the faces of the $k$-cell. So, a 0-dimensional cell is a point, a 1-cell is a line segment, a 2-cell is a polygon, a 3-cell is a solid polyhedron with polygons, edges, and vertices as faces. Thus, the endpoints of a 1-cell are 0-cells, the boundary of a 2-cell consists of 1-cells, etc. A finite number of cells glued together is said to form a cell complex $K$ as long as the following two conditions hold: if $\sigma$ and $\tau$ are $k$-cells in $K$, then all $(k-1)$-cells of $\sigma$ and $\tau$ are elements of $K$, and Int$(\sigma) \cap \mbox{ Int}(\tau) = \emptyset$, where Int(A) denotes the interior of A. The dimension $\kappa$ of a cell complex is defined by the dimension of its highest-dimensional cells.

A complex $K$ is much more than a set of points or a tessellation: it is a layered construction equipped with a structure of cells of various dimensions. To track all the cells from a complex and how they are glued together, we must also consider the directions of any edges glued together. A complex $K$ of dimension $\kappa=2$ is said to be oriented if each 1-cell is given a direction (from initial point to terminal point) and each 2-cell is given a direction (clockwise or counterclockwise). Note that the choices of directions for 1-cells and 2-cells are {\it a priori} arbitrary. For oriented complexes $K$ of dimension $\kappa$, a $k$-chain $C_k$ is defined as
\begin{equation}
  C_k = \sum_i a_i \, \sigma^i_k
\end{equation} where $a_i \in \mathbb{Z}$ and $\sigma^i_k$ is a $k$-cell 
($\forall i \in \mathbb{N}$). The group of all $k$-chains ${\cal C}_k$ 
($k=1 ,\cdots,\kappa$) in $K$ is an abelian group. Two $k$-cells are adjacent 
if they share a $(k-1)$-cell. Thus, a boundary operator 
\[ \partial_k : {\cal C}_k\rightarrow {\cal C}_{k-1} \,  \] is introduced in such a way that, for instance, the boundary of an oriented 2-cell is the chain formed by the 1-cells on its border, with a positive sign if the orientation of an edge is consistent with the direction of the 2-cell, and with a negative sign otherwise. By construction, the boundary operator satisfies
\[ \partial_k \circ \partial_{k+1} \rightarrow 0_{k+1,k-1} \, , \]
where $0$ denotes the trivial group, that is, the constant map sending every element of ${\cal C}_{k+1}$ to the group identity in ${\cal C}_{k-1}$. One can show that $\partial_k$ is a homomorphism for any complex $K$. Stating that ``the boundary of a  boundary is trivial'' is equivalent to stating that $\mathrm{im} (\partial_{k+1}) \subseteq \ker(\partial_k)$, where $\mathrm{im} (\partial_{k+1})$ designates the image of the boundary operator and $\ker(\partial_k)$ its kernel. The elements of the group ${\cal B}_k(K) = \mathrm{im} (\partial_{k+1})$ are called $k$-boundaries, and the elements of ${\cal Z}_k(K) = \ker(\partial_k)$ are called $k$-cycles. Note that a $k$-cycle $C_k$ in ${\cal C}_k$ is such that $\partial_k (C_k) = 0$, that is, a $k$-cycle has no boundary. Among the $k$-cycles, one can distinguish those that are the boundaries of some ($k+1$)-cells (they
are thus $k$-boundaries), and they constitute the group ${\cal B}_k$.
 
The groups ${\cal C}_k$, ${\cal Z}_k$, and ${\cal B}_k$ depend on the particular complex that is built. To excerpt the main properties of the topology of the underlying cloud of points, an equivalence relation is introduced on chains. 

\begin{theor}
Let $C_k^{\alpha_1}$ and $C_k^{\alpha_2}$ be two $k$-chains $\in$ ${\cal C}_k$. 
They are said to be {\tt homologous} --- noted $C_k^{\alpha_1} \sim 
C_k^{\alpha_2}$ --- if and only if there exists a boundary of a ($k+1$)-chain 
$C_{k+1}$ such that $\partial_{k+1}(C_{k+1})=C_k^{\alpha_1} - C_k^{\alpha_2}$.
\end{theor}

Since the chain group ${\cal C}_k$ is abelian, all its subgroups are normal. It 
is thus possible to introduce the quotient group
\[
  {\cal H}_k  := \ker(\partial_k) / \mathrm{im} (\partial_{k+1})
  = {\cal Z}_k / {\cal B}_k  \, , 
\]
called the $k$th homology group of $K$ made of homology classes over cycles. We denote a homology group in terms of its generators, that is, as ${\cal H}_k = \left[ \displaystyle g_1, ..., g_q \right]$, with $q \in \mathbb{N}$. The cardinal $q$ of ${\cal H}_k$ corresponds to the $k$th Betti number $\beta_k$. Notice 
that the $k$-generators $g_i$ of a homology group ${\cal H}_k$ are homologically independent, that is, they cannot be deformed into each other by a continuous transformation (an isotopy). 
The homology groups thus cancel out unnecessary information for characterizing the underlying object. In each dimension, ${\cal H}_k(K)$ gives a piece of information related to the properties determined by that dimension. Thus, (i) the group ${\cal H}_0(K)$ measures the connectivity of the complex, and the rank of ${\beta}_0$ refers to the number of connected components, (ii) the group ${\cal H}_1(K)$ identifies non-trivial loops around the complex through the 1-generators, (iii) the group ${\cal H}_2(K)$ identifies the 2-generators of ${\cal H}_2$ which identify the enclosed cavities of the complex $K$. 

The number of $k$-generators (the Betti numbers $\beta_k$) do not depend on the particular complex that is built, and allow to easily distinguish some topologically non equivalent manifolds. 
Even if the generators are necessarily written in terms of the labelled cells of the particular complex, identifying and locating them in the complex is highly relevant, since their relative intertwining offers information about how the underlying object is structured. 

It is however important to remark that there is much more information contained in a cell complex than that encoded by the Betti numbers and the explicit generators of the homology groups. Additional properties can be extracted if all the cells in the complex are oriented in the same way, or, in other words, if the complex is uniformly oriented. \cite{Sci99,Sci01} Two cells are said to have the same orientation if they are oriented so that the common edges are canceled when summing the borders of the two cells. Assigning a uniform orientation to a complex is simple if the underlying manifold is Haussdorff. A space is Hausdorff if, 
for any two distinct points in space, there exist two open sets (one for each 
point) which do not intersect. Let us assume that this is the case. 
One can hence start by assigning the same orientation to all the 2-cells of a 2-complex $K$ ($\kappa=2$). This is done by choosing an orientation for one 2-cell (e.g. clockwise) and propagating it across the rest of the 2-cells, till they are all oriented in the same manner. This yields a uniformly oriented complex $K^u$. 

Now, let $\Gamma=\sum_i \sigma^i_2$ where $\sigma^i_2$ are all the 2-cells in $K^u$. Let us write $$\partial(\Gamma)= \underbrace{\sum_j a_j \sigma^j_1 }_\text{boundaries} + \underbrace{\sum_j b_j \sigma^j_1}_\text{torsions},$$ 
\noindent
where $a_j=\pm 1$, $b_j\ne \pm 1$, and $\sigma^j_1$ are 1-cells in $K^u$. The first term refers to the boundary of the manifold underlying the uniformly oriented complex, and the second term to some torsion elements. If the first term is null, the manifold approximated by the complex is said to have no boundary, as for a sphere, a torus or a Klein bottle. On the other hand, the underlying manifold is said to be non-orientable if the second term is non zero, that is, if $\partial \Gamma(K^u)$ contains torsion elements.  


Table \ref{bettinum} lists some of the topological properties (Betti numbers $\beta_i$, existence of boundaries $b$ or torsions elements $t$) for some common surfaces. The important message here is that a finer description can be obtained when one is working 
with a uniformly oriented complex.\citep{Cha21}

\begin{table}[ht]
   \caption{Betti numbers $\beta_k$, existence of torsion elements $t$ or boundaries $b$ for some 2-dimensional manifolds.  }
  \label{bettinum}
  \centering
  \begin{tabular}{lcccccccc}
	  \hline \hline
	  \\[-0.3cm]
	  &  $\beta_0$ & $\beta_1$ & $\beta_2$ & $t$ &$b$ \\[0.1cm]
	  \hline
	  \\[-0.3cm]
	  Disk             &1 &    0 &  0 & No & Yes \\[0.1cm]
	  Cylinder        &   1 &    1 &  0 & No  & Yes \\[0.1cm]
	  M\"obius band  &   1 &    1 &  0 &  Yes  & Yes \\[0.1cm]
	  Torus           &   1 &    2 &  1 & No  & No \\[0.1cm]
	  Genus-$g$ torus &   1 & $2g$ &  1 & No & No \\[0.1cm]
	  Klein bottle 	  &   1 &    1 &  0 & Yes  & No \\[0.1cm]
	  \hline \hline
  \end{tabular}
\end{table}

Branched manifolds are mathematical objects which are not Hausdorff. 
In order to endow the corresponding complex with a uniform orientation that is compatible with the flow, 
one must decompose the complex in sub-units, as a template must be decomposed into strips. This decomposition leads us to the concept of {\it templex} as introduced in the next section.


\subsection{Templex}
\label{sectemplex}

State space is a ubiquitous concept that is especially relevant in chaos and 
nonlinear dynamics.\cite{Nol10} A cloud of points in an attractor is plotted in 
space, but each point in this particular space represents a state at a given
time. When one attempts to describe such sets of points with a cell complex, an 
important piece of information is missing. This information concerns the flow. 
Templates are in fact constructed taking the flow information into account. In 
order to bridge the gap between cell complexes and templates, we need a 
mathematical object going beyond a cell complex. According to the flow, from a 
given $\kappa$-cell of the complex, it is only possible to go towards a few 
other $\kappa$-cells. These connections can be expressed in terms of a directed graph or digraph, whose nodes denote the highest dimensional cells of a complex, and whose directed edges denote allowed transitions between them.  This combination of a cell complex and a digraph will be useful to encode all the information contained in a template, and will be therefore baptized with the term 
``templex''.

\vspace{0.3 cm}
\paragraph*{\underline{Step 0: BraMAH complex}}

In order to construct a templex, let us start from a BraMAH complex obtained 
from a set of points associated with an attractor in a state space. The 
orientation of the cells at this stage is not necessarily uniform: any 
orientation is valid. Let us assume that our attractor lies in a branched 
2-manifold, so that the dimension of the BraMAH complex is $\kappa = 2$. We 
will use the Lorenz attractor as an example.

\vspace{0.3 cm}
\paragraph*{\underline{Step 1: Locating the joining locus}}

The particularity of a branched $2$-manifold as underlying a chaotic attractor 
is to present a joining chart at which more than two 2-cells share one 1-cell. The first step is to locate the 1-cells shared by at least three 2-cells: the 1-chain of these 1-cells is called the {\it joining locus}. 

\vspace{0.3cm}
\paragraph*{\underline{Step 2: Re-orienting the complex}}

\begin{figure}[htbp]
  \centering
  \begin{tabular}{c}
    \includegraphics[width=0.44\textwidth]{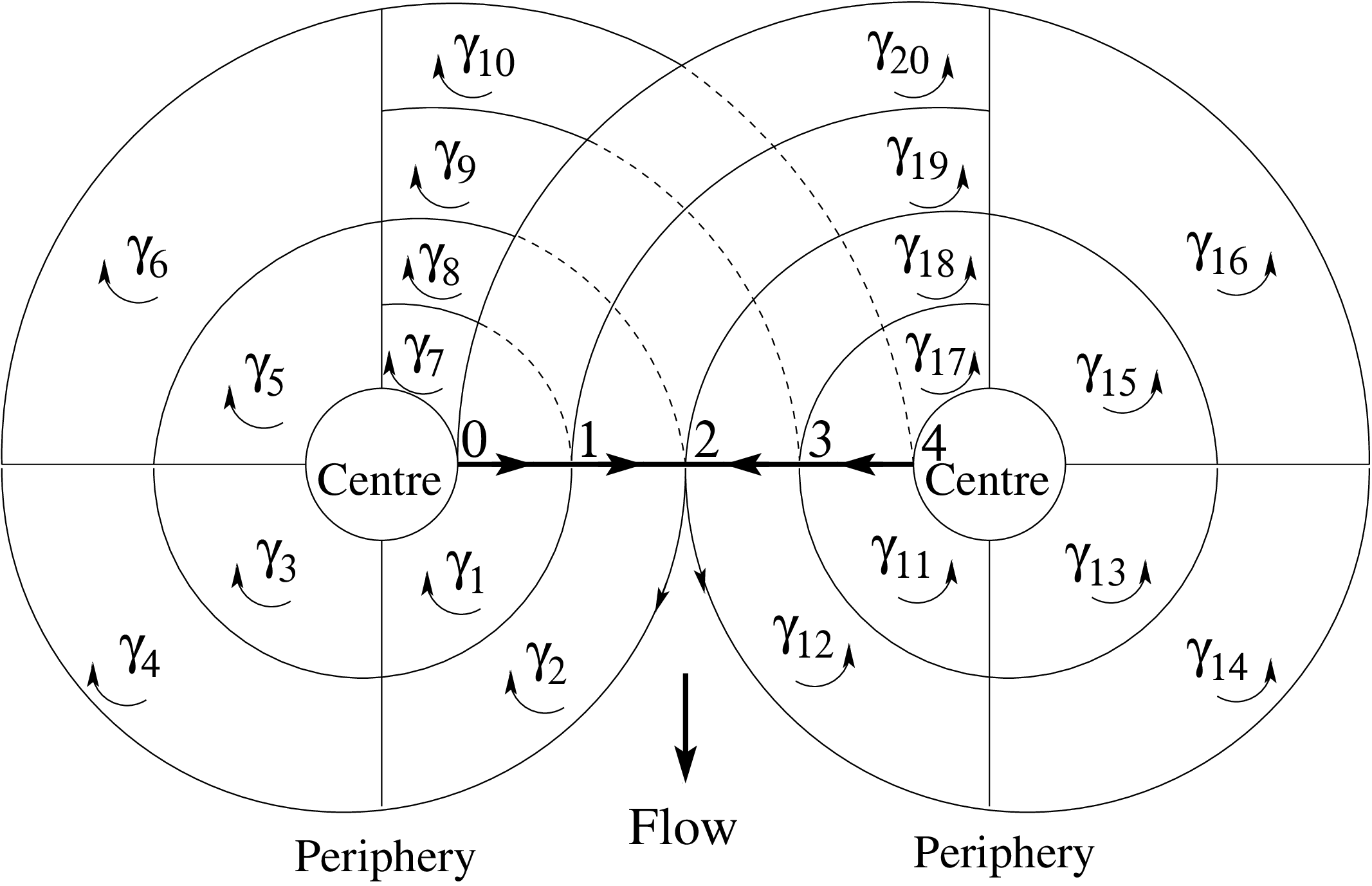}\\[0.2cm]
	  (a) Flow oriented complex $K_1({\rm L})$ \\[0.2cm]
    \includegraphics[width=0.44\textwidth]{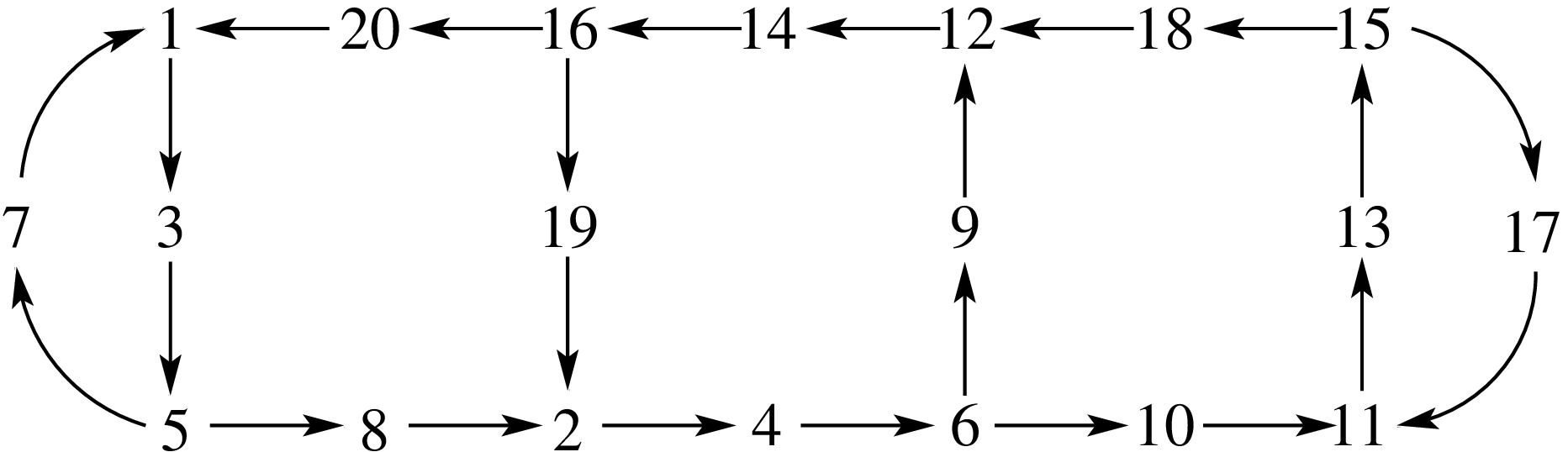} \\[0.2cm]
	  (b) Associated digraph $G_1({\rm L})$  \\[0.2cm]
    \includegraphics[width=0.44\textwidth]{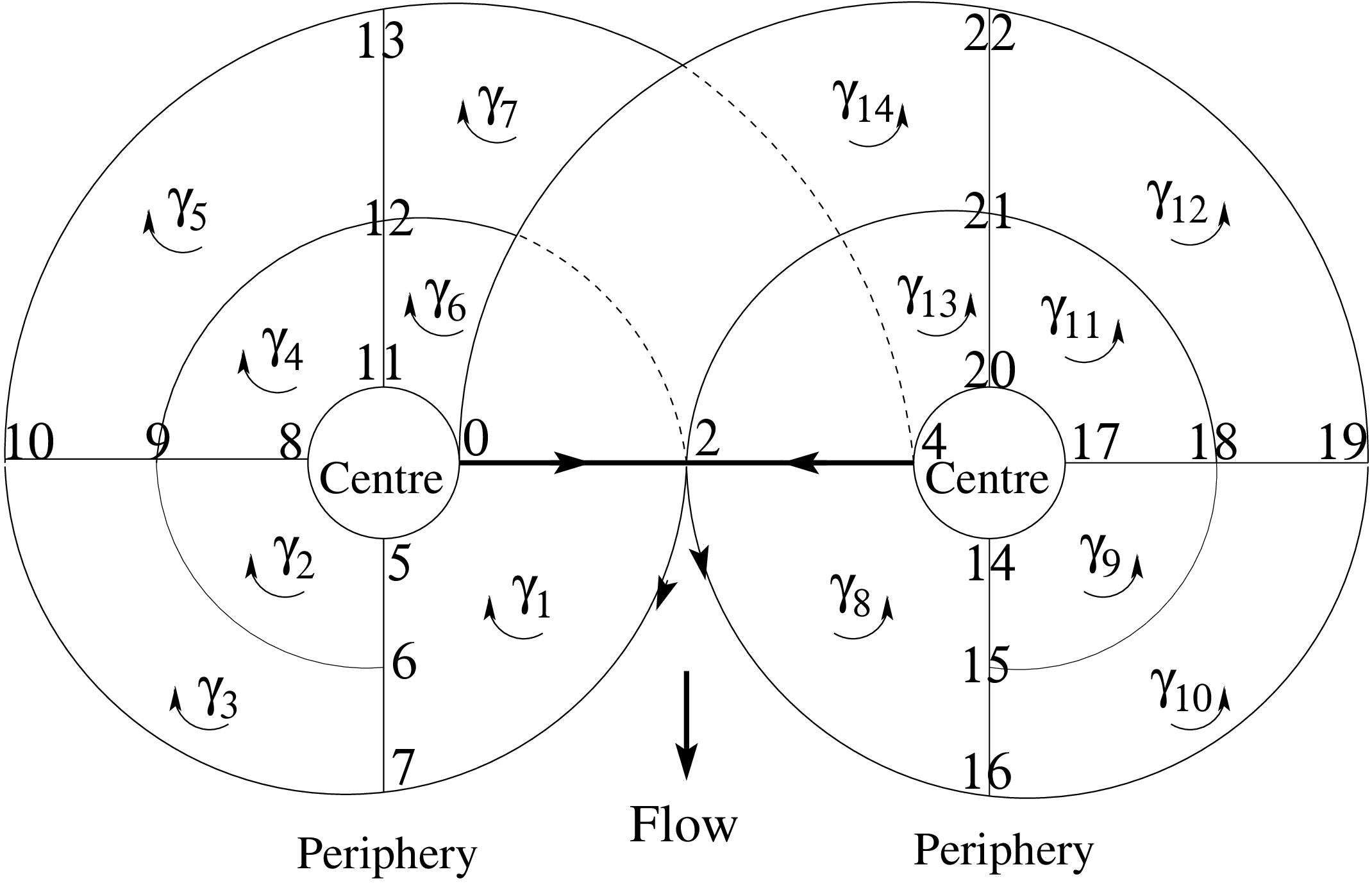} \\
	  (c) Generating complex $K_2({\rm L})$ \\[0.2cm]
    \includegraphics[width=0.38\textwidth]{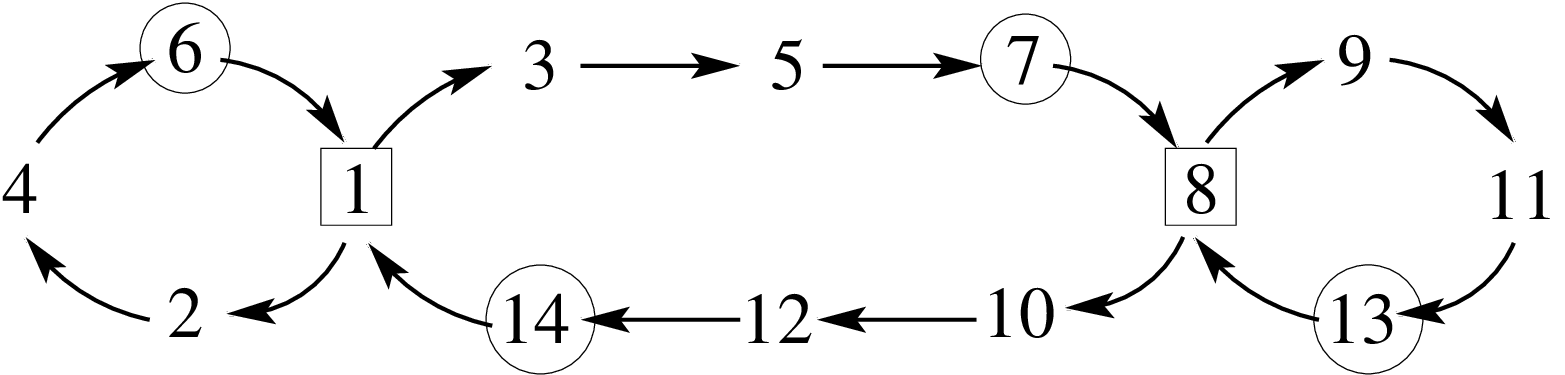} \\[0.2cm]
	  (d) Generating digraph $G_2({\rm L})$  \\[0.2cm]	  
    \includegraphics[width=0.40\textwidth]{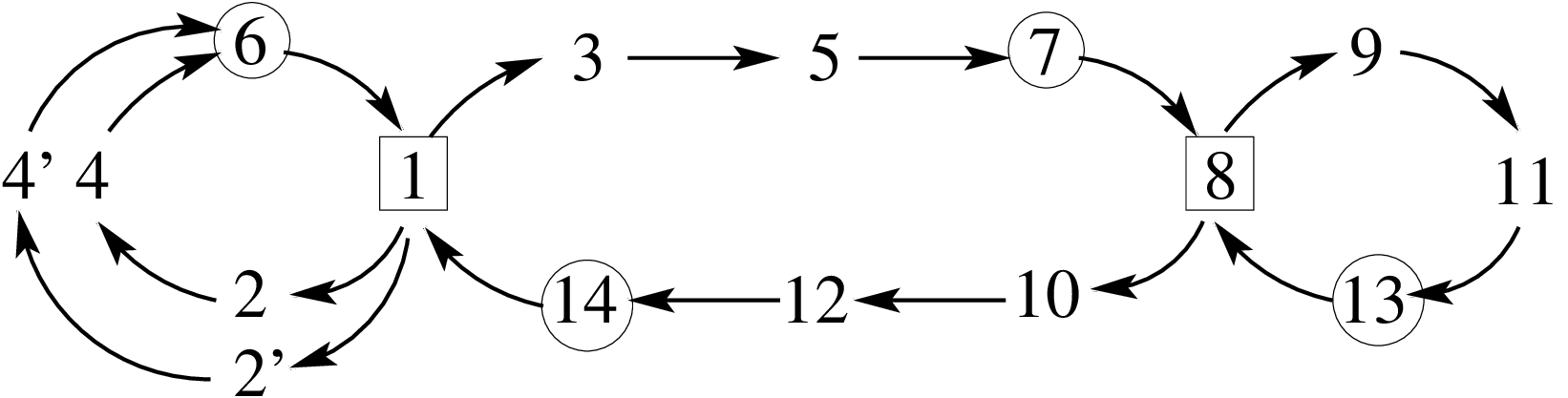} \\[0.2cm]
	  (e) Generating digraph $G'_2({\rm L})$  \\[-0.2cm]	  
  \end{tabular}
	\caption{(a-b) Templex $T_1({\rm L})=(K_1({\rm L}),G_1({\rm L}))$ and (c-d) generating 
	templex $T_2({\rm L})=(K_2({\rm L}),G_2({\rm L}))$ for the Lorenz attractor. In (e), digraph 
$G_2'({\rm L})$ corresponding to another generating complex $K_2'({\rm L})$
--- not shown --- for the Lorenz attractor.
Ingoing and outgoing nodes are squared and circled, respectively.
  }
  \label{complorun12}
\end{figure}

A 2-cell with one 1-cell at the periphery of the complex is arbitrarily chosen: this 1-cell is oriented according to the flow. For the complex in Fig.\ \ref{complorun12}(a), say, for instance, that the 2-cell $\gamma_2$ is chosen: the peripheral 1-cell is thus oriented along the flow and, consequently, the 2-cell is 
clockwise. 

The other 2-cells are oriented by a simple propagation of this orientation up to the joining locus. Indeed, one of the specificities of a templex is that {\bf 
the 2-cell orientation cannot be propagated across a joining locus}. If there are still 2-cells which have not yet been re-oriented (because they are located across a joining locus with respect to the initially chosen 2-cell), the procedure is repeated from another arbitrarily chosen 2-cell not yet re-oriented, and so on, up to the stage where all 2-cells are re-oriented. A complex that is oriented according to this rule will be said to be a {\bf flow-oriented complex}. 

In our example, the flow-oriented complex $K_1({\rm L})$ drawn in Fig.\ 
\ref{complorun12}(a) for the Lorenz attractor is obtained from two starting 
2-cells (for instance, $\gamma_2$ and $\gamma_{12}$).

\vspace{0.3cm}
\paragraph*{\underline{Step 3: Constructing a digraph}}

A digraph related to the way 2-cells are visited by the flow is constructed 
as follows. Two nodes $i$ and $j$ are connected as $i \rightarrow j$ when the 
flow provides a path from $\gamma_i$ to $\gamma_j$. This naturally leads to the 
introduction of the notion of {\it templex}. 

\begin{defi}
A {\bf templex} $T \equiv (K,G)$ is made of a complex $K$ of dimension 
dim$(K)=\kappa$ and a digraph $G=(N,E)$ whose underlying space is a branched 
$\kappa$-manifold associated with a dynamical system, such that (i) the nodes 
$N$ are the $\kappa$-cells of $K$ and (ii) the edges $E$ are the connections between the $\kappa$-cells allowed by the flow. 
\end{defi}

An example of templex $T_1({\rm L})= (K_1({\rm L}),G_1({\rm L}))$ for the Lorenz attractor is drawn in 
Figs.\ \ref{complorun12}(a) and \ref{complorun12}(b).

\vspace{0.3cm}
\paragraph*{\underline{Step 4: Splitting the joining locus}}

Let us designate as the {\it outgoing} 2-cells, the 2-cells which are visited by the flow just after crossing one of the joining 1-cells belonging to the joining locus. Each joining 1-cell must be oriented according to the outgoing 2-cell from which it is an edge. If there is a change in direction within a 
connected set of joining 1-cells --- as the 0-cell $\langle 2 \rangle$ in Fig.\ 
\ref{complorun12}(a) --- then there is a {\it splitting} 0-cell which divides
this connected set into two different connected sets of joining 1-cells. Once
this is completed,
each connected set of 1-cells sharing the same direction corresponds to one
component of the Poincaré section. Each component is denoted by $J_i$ where $i \in \mathbb{N}$. 

In the flow-oriented complex $K_1$ of Fig.\ \ref{complorun12}(a), the joining locus is split in two chains $J_1 = \langle 0, 1 \rangle + \langle 1, 2 \rangle$ and $J_2 = - \langle 2, 3 \rangle - \langle 3, 4 \rangle$. 
Here, $\langle 0, 1 \rangle$ denotes the 1-cell starting in the 0-cell $\langle 0 \rangle$ and finishing in the 0-cell $\langle 1 \rangle$. Further details concerning notation are given in appendix A. 
The joining locus of the complex $K_1$ has thus two different components, $J_1$ and $J_2$. The 0-cell $\langle 2 \rangle$ is a {\it splitting} 0-cell.

\vspace{0.3cm}
\paragraph*{\underline{Step 5: Removing superfluous joining 0-cells}}

For each chain $J_i$ of the joining locus, let us remove the 0-cells that are 
not its boundaries. This induces a simplified flow-oriented complex, with a 
minimal structure for the joining locus. For instance, the two joining 1-cells
$\langle 0,1 \rangle$ and $\langle 1,2 \rangle$ are merged into a single one ($\langle 1,2 \rangle$), 
by removing the 0-cell $\langle 1 \rangle$. These two 1-cells are merged, and 
so the 2-cells attached to it. After this simplification, 
the complex has only one outgoing 2-cell per component of the joining locus 
[Fig.\ \ref{complorun12}(c)]. This so-reduced flow-oriented complex and its 
associated digraph $G_2({\rm L})$ is what we shall call a {\it generating templex}. 
Here ``generating'' refers to the generating partition induced by the 
first-return map. In a generating templex, the number of components of the 
joining locus is equal to $N_{\rm loc}$, each joining locus being made of a 
single 1-cell. There are $N_{\rm loc}$ outgoing 2-cells, one per component 
of the Poincaré section. We have thus $N_{\rm loc} = N_{\rm c}$, a number which
cannot be reduced further.

\begin{defi}
A {\bf generating templex} $T^{\rm g} \equiv (K^{\rm g},G^{\rm g})$ is a templex composed by a flow-oriented complex $K^{\rm g}$ associated with a digraph $G^{\rm g}$ such that there is only one outgoing 2-cell per component of the joining locus.  
\end{defi}



By propagating this removal of 0-cells to the higher-dimensional 1-cells, the 
2-cells $\gamma_1$ and $\gamma_2$ are merged in a single one, and similarly for the pair $\gamma_7$-$\gamma_8$, $\gamma_9$-$\gamma_{10}$, $\gamma_{11}$-$\gamma_{12}$, $\gamma_{17}$-$\gamma_{18}$, and $\gamma_{19}$-$\gamma_{20}$. This leads to the complex $K_2({\rm L})$ drawn in Fig.\ \ref{complorun12}(c). The resulting complex $K_2({\rm L})$ has a joining locus with two components as expected, since the Poincaré section of the Lorenz attractor has two components. Trajectories in the attractor can cross the joining locus 
from four ingoing 2-cells, namely $\gamma_6$, $\gamma_7$, $\gamma_{13}$, and 
$\gamma_{14}$. Each joining locus has two ingoing 2-cells and one outgoing 
2-cell. For instance, $J_1$ has $\gamma_6$ and $\gamma_{14}$ as 
ingoing 2-cells, and $\gamma_1$ as outgoing 2-cell. Notice that the total number of ingoing 2-cells is equal to the number $N_{\rm s}$ of strips in the corresponding template (4 for the Lorenz template as discussed in Section \ref{loratt}). 

A generating templex is not unique and the number of 2-cells (or nodes) that do 
not share a joining 1-cell remains arbitrary. 

\vspace{0.3cm}
\paragraph*{\underline{Step 6: Computing cycles of the generating digraph }}

Let us now compute the cycles of the generating digraph [Fig.\ 
\ref{complorun12}(d)]. Due to the construction of the digraph, it is efficient
to compute the cycles from the outgoing nodes. From the generating
digraph $G_2 ({\rm L})$, there are the three cycles 
\[ c_1 \equiv \underline{1} \rightarrow 2 \rightarrow 4 \rightarrow 6 
  \rightarrow \underline{1} \]
\[ c_2 \equiv \underline{8} \rightarrow 9 \rightarrow 11 \rightarrow 13 
  \rightarrow \underline{8} \]
and
\[ c_3 \equiv \underline{1} \rightarrow 3 \rightarrow 5 \rightarrow 7 
  \rightarrow \underline{8} \rightarrow 10 \rightarrow 12 \rightarrow 14 
  \rightarrow \underline{1} 
  \]
Let us imagine another generating complex $K_2' ({\rm L})$ with more numerous
2-cells than $K_2 ({\rm L})$, the 2-cells $\gamma_2$ and $\gamma_4$ being split
in two, leading to the addition of $\gamma_2'$ and $\gamma_4'$, respectively.
The corresponding digraph $G_2' ({\rm L})$ is drawn in Fig.\ 
\ref{complorun12}(e). There is the additional cycle
\[ c_1' \equiv \underline{1} \rightarrow 2' \rightarrow 4' \rightarrow 6 
  \rightarrow \underline{1} \, . \]
By construction, the cycles $c_1$ and $c_1'$ are equivalent because their 2-chains visit the same ingoing and outgoing 2-cells. Note that in terms of template, they would belong to the same strip. 
%
Consequently, among the set of cycles of the digraph $G$, if there 
is more than one cycle visiting a given ingoing and a given outgoing nodes, only one is retained. From the digraph $G_2' ({\rm L})$, the 
cycle $c_1'$ (for instance) would be discarded.

Let us now introduce the order $p$ of cycle as the number of ingoing nodes. For the digraph $G_2 ({\rm L})$, there two order-1 cycle ($c_1$ and 
$c_2$) and one order-2 cycle ($c_3$). The latter is in fact degenerated and 
shall be considered as the union of two weak cycles, namely
\[ c_{3_1} \equiv \underline{1} \rightarrow 3 \rightarrow 5 \rightarrow 7 
  \rightarrow \underline{8} \]
and
\[ c_{3_2} \equiv \underline{8} \rightarrow 10 \rightarrow 12 
  \rightarrow 14 \rightarrow \underline{1} \, . \]
In the Lorenz attractor, the weak cycles $c_{3_1}$ and $c_{3_2}$ are
the images of each other under the action of the rotation symmetry.
They are weak as boundaries are weak when they have to be travelled more than 
once to become actual boundaries. 

With each cycle of the generating digraph $G^{\rm g}$, there is an associated
chain of 2-cells which forms a sub-templex.

\begin{defi}
Each cycle from the generating digraph $G^{\rm g}$ associated with its 
subcomplex forms a subtemplex that is called {\bf generatex} $\mathcal{G}$. A 
generatex is said to be of order $p$ with $p \in \mathbb{N}$, $p \ge 1$, if 
its cycle has $p$ distinct ingoing nodes. 
\end{defi}
A generatex is said to be {\it simple} if $p = 1$, and {\it degenerated} if 
$p > 1$. 

The union of all equivalent sub-templexes (associated with equivalent cycles 
from the generating digraph) is a unit which plays a fundamental role in the 
characterization of branched manifolds. As we only retain a single representative for each group of equivalent cycles, we only save the representative generatexes. 

%

\begin{defi}
A {\bf stripex} $\mathcal{S}_i$ is a subtemplex associated to a cycle or to a weak cycle of a generatex.
\end{defi}

By construction, each stripex is a subtemplex of $T^{\rm g}$. 
A stripex is the analog of a strip from a template. Each strip is closely 
related to the structure of the first-return map to the Poincaré section.

In $T_2({\rm L})$, for instance, there are four stripexes: $\mathcal{S}_1$ associated to $c_1$, $\mathcal{S}_2$ associated to $c_2$, $\mathcal{S}_3$ associated to $c_{3_1}$ and $\mathcal{S}_4$ associated to $c_{3_2}$.

\vspace{0.3cm}
\paragraph*{\underline{Step 7: Computing local twists}}

The two free edges of the complex in stripex $\mathcal{S}_i$ are defined as the two disconnected 1-chains of the associated sub-complex $K_i$ that result from applying the boundary operator $\partial_2$ to the sum of all the cells in $K_i$ but the 
1-cells from the joining locus. 

\begin{defi}
A stripex $\mathcal{S}_i = (K_i,G_i)$ is said to have a {\bf local twist} if 
the free edges of $K_i$ change their relative positions with respect to the 
orientation from the center to the periphery.
\end{defi}

Note that in the correspondence with templates, uneven local torsions in a strip correspond to a local twist.  Strips with no local torsion or with even parity local torsions correspond to non twisted stripexes. 

In $T_2({\rm L})$, the 2-cells in stripex $\mathcal{S}_3$ $(\gamma_1, \gamma_3, \gamma_5, \gamma_7)$ and in $\mathcal{S}_4$ $(\gamma_8, \gamma_{10},\gamma_{12}, \gamma_{14})$ have a local twist. For each of them, the free edge
starting in the center ends at the periphery, and {\it vice versa}.

\vspace{0.3cm}
\paragraph*{\underline{Step 8: Drawing the templex}}

The standard representation of the different strips in a template introduces crossings to indicate local torsions and a certain order to organize the position of the strips, namely, which one is on top of the other and how they permute.  

\begin{figure}[ht]
  \centering
  \begin{tabular}{cc}
    \includegraphics[width=0.23\textwidth]{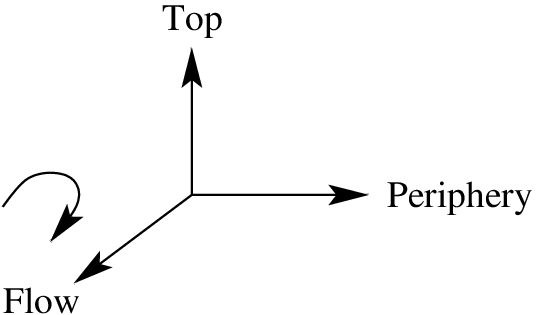} &
    \includegraphics[width=0.155\textwidth]{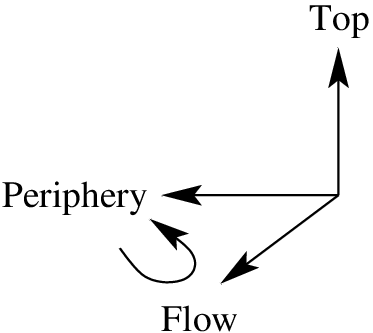} \\
    {\small (a) Clockwise flow} & {\small (b) Anti-clockwise flow} \\[-0.2cm]
  \end{tabular}
  \caption{Orientation of the normal to a 2-cell to define relative top (and 
bottom). This is required for determining the order with which 2-cells are 
joined at a joining locus (see below).}
  \label{trihed}
\end{figure}

The graphical representation used in Figure 4 is a sort of `semi-planar' diagram that keeps the shape of the butterfly for reference, giving an idea of how the complex is attached to the Lorenz attractor. Drawing a complex in this manner requires knowing which 2-cell is at the top of the other at the joining locus. For instance, when cells $\gamma_6$ and $\gamma_{14}$ are considered, one should be able to determine which one is at the top of the other. Since we are in a three-dimensional space, it is important to locally orientate the direction of the top. To do this, we use the orientation of the flow and of the joining 1-cell. For a clockwise flow (counter-clockwise), we use the right (left) trihedron with the triplet flow-periphery-top as sketched in Fig. \ref{trihed}(a) [Fig.\ \ref{trihed}(b)].  According to this, in $T_2({\rm L})$, $\gamma_{14}$ should be drawn at the top of $\gamma_6$ and $\gamma_{13}$ at the top of $\gamma_7$.

\vspace{0.3cm}
\paragraph*{\underline{Correspondence with templates}}

To sum up, templexes enable to bridge the gap in the description of chaotic 
attractors offered by cell complexes and templates. The topological properties 
of an attractor are encoded by a generating templex. 
For a branched 2-manifold, the description provided by the generating templex is organized in terms of a number of layered properties:

\begin{enumerate}

\item[$\blacksquare$] Splitting 0-cells are a particular case of critical points, appearing between two different components of the Poincaré sections. 

\item[$\blacksquare$] Joining 1-cells in the generating complex are the analogs of the components of the Poincaré section.


\item[$\blacksquare$] Joining 2-cells are the analogs of the joining charts in a template. 

\item[$\blacksquare$] Stripexes are the analog of strips in a template: each stripex corresponds to a period-1 orbit in the template. 


\end{enumerate}

\section{Characterization of 3D chaotic attractors}
\label{bramako}

We partly addressed the case of the Lorenz attractor when introducing the concept of templex. Let us now describe its template to exhibit its equivalence with the constructed templex. The cases of the R\"ossler attractor, and the Burke-and-Shaw attractor are then treated. 

\subsection{The Lorenz attractor}
\label{loratt}

\noindent
{\underline{Template}}
\vspace{0.3 cm}

\begin{figure}[ht]
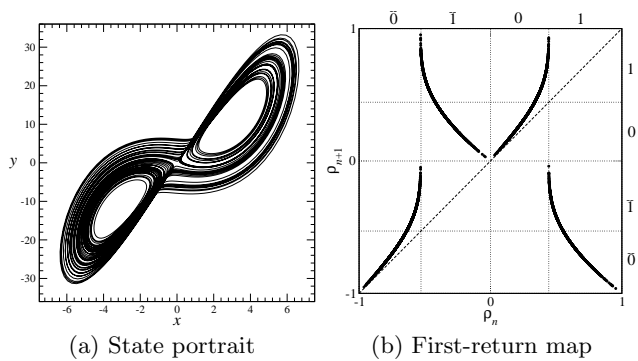

  \centering
  \begin{tabular}{cc}
    \includegraphics[width=0.2280\textwidth]{loratt63.eps} &
    \includegraphics[width=0.23\textwidth]{lor2map.eps} \\
    (a) State portrait & (b) First-return map \\[-0.2cm]
  \end{tabular}
  \caption{Chaotic attractor produced by the Lorenz system (\ref{loreq63}) when the symbolic dynamics is nearly complete. Parameter values: $R=28$, $\sigma = 10$, and $b = \frac{8}{3}$.}
  \label{loratt63}
\end{figure}

\begin{figure}[ht]
  \centering
  \begin{tabular}{c}
	 \includegraphics[width=0.33\textwidth]{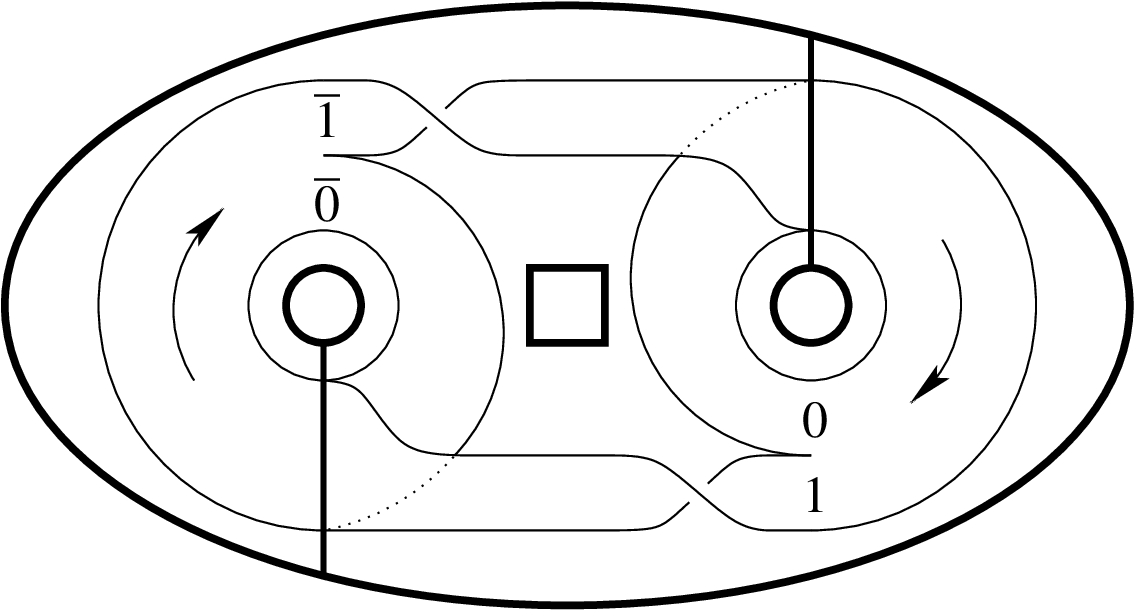}  \\[0.1cm]
         (a) Template and genus-3 bounding torus \\[0.2cm]
	 \includegraphics[width=0.33\textwidth]{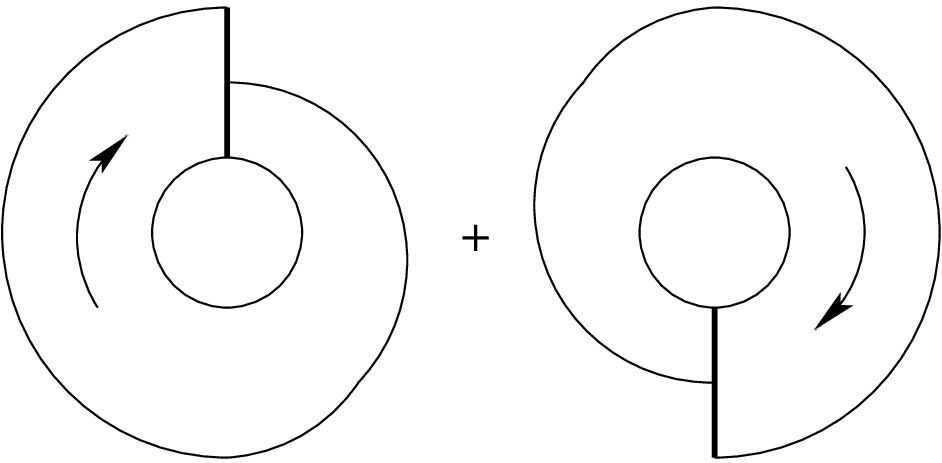} \\
	 (b) Strips 0 and $\0$ \\[0.2cm]
	 \includegraphics[width=0.33\textwidth]{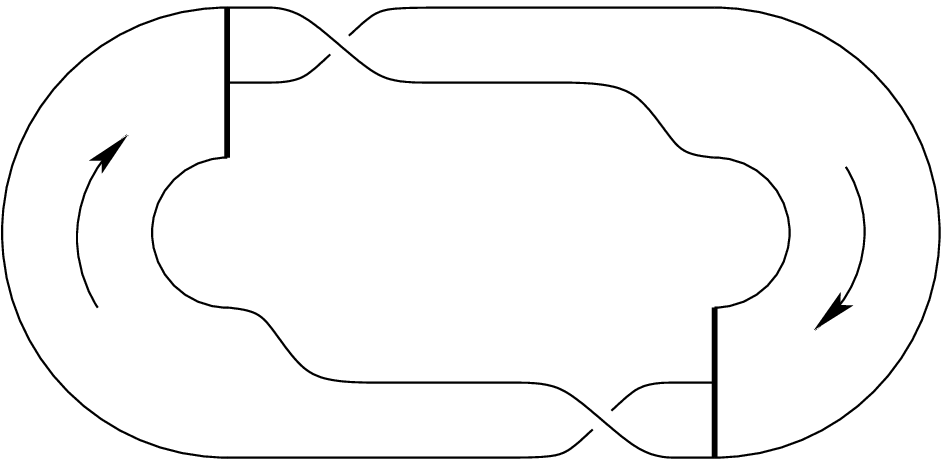} \\[0.1cm]
	 (c) Union of strips $1$ and $\1$ \\[-0.2cm]
  \end{tabular}
\caption{Template and genus-3 bounding torus (thick lines) (a) for the Lorenz attractor and its decomposition into strips. The union of strips $1$ and $\1$ (c) is degenerated since it crosses twice the joining 1-line. }
  \label{bramalor}
\end{figure}

The Lorenz attractor [Fig.\ \ref{loratt63}(a)] produced by the Lorenz system\cite{Lor63}
\begin{equation}
  \label{loreq63}
  \left\{
    \begin{array}{l}
      \dot{x} = - \sigma x + \sigma y \\[0.1cm]
      \dot{y} = Rx -y -xz \\[0.1cm]
      \dot{z} = -b z +xy 
    \end{array}
  \right.
\end{equation}
is bounded by a genus-3 torus which has three holes, two of the focus type which are associated with the center of each wings, and one, located in the rotation axis, which is a saddle type.\cite{Let05a} The first two ones can be easily detected because there are circled by the attractor. This is not so obvious for the hole of the saddle type, as drawn in Fig.\ \ref{bramalor}(a), because it is bounded by trajectories issued from each of the two wings, that is, from different components of the Poincaré section. 

Let us first show how the Lorenz template is unveiled. The first step is to start from a view of the Lorenz attractor in the $x$-$y$ plane and not in the $x$-$z$ plane or in the $y$-$z$ plane as commonly done. The template for the Lorenz attractor is drawn in Fig.\ \ref{bramalor}(a). As suggested by the first-return map, it should be decomposed into four different strips. Two are the strips 0 and $\0$ [Fig.\ \ref{bramalor}(a)] and one is the strip corresponding to the union of strips 1 and $\1$ [Fig.\ \ref{bramalor}(b)]. 


\begin{figure}[ht]
  \centering
  \includegraphics[width=0.43\textwidth]{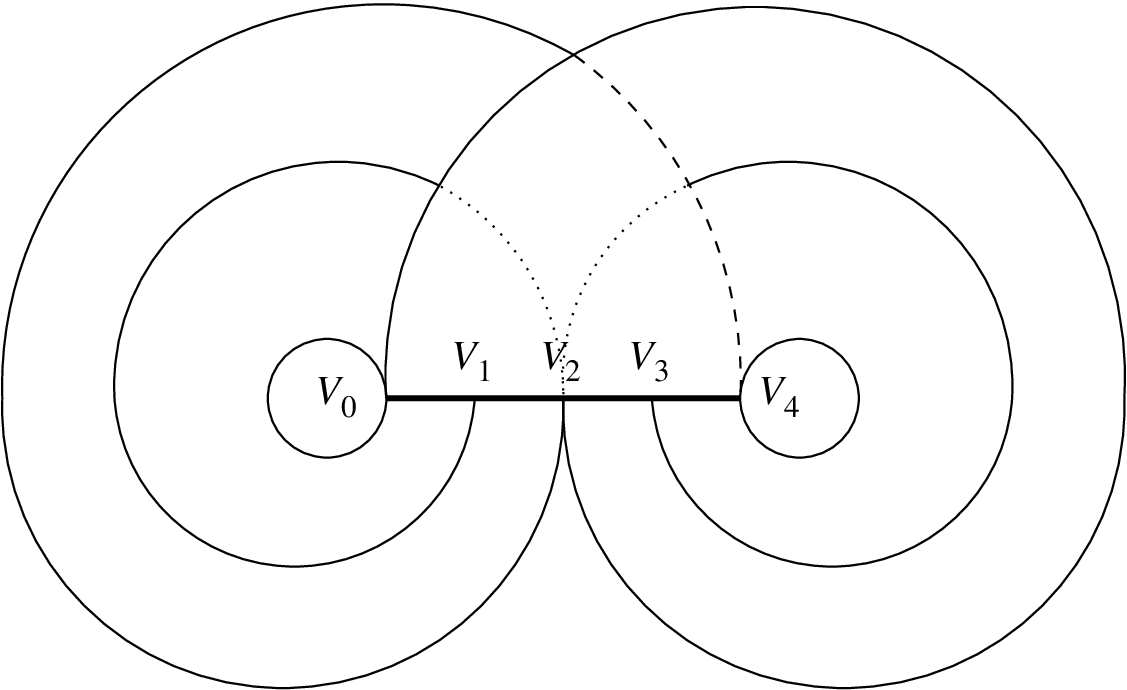} \\[-0.2cm]
  \caption{Decomposition of the template for the Lorenz attractor in four 
strips from an $x$-$z$ view.}
  \label{bramalorxz}
\end{figure}

There is another possibility to represent the template of the Lorenz attractor: it results mostly from a projection of the attractor in the $x$-$z$ plane or in the $y$-$z$ plane (Fig.\ref{bramalorxz}). In that particular representation, the Poincaré section may appear as being formed from a single component but there are a few characteristics which imposes to split it in two different components. Moreover, it is needed to orient the Poincaré section from the center of the attractor to its periphery. Typically, the center of an attractor is associated with the focus point around which trajectories circle. Consequently, the center of the attractor is different for each wing: the center of the left wing is at the left of the Poincaré section and the center of the right wing is at the right of the Poincaré section. This means that the Poincaré section has necessarily two components which are oriented in opposed directions. Thus the point 
$V_3$ in Fig.\ \ref{bramalorxz} necessarily splits the Poincaré section in two 
components. This points results from the action of the rotation axis --- the $z$-axis --- at which the transverse flow is necessarily null.\cite{Let01} The point $V_3$ thus splits the template issued from the joining line in two parts, one being associated with the left wing, the other one with the right wing.

\begin{figure}[ht]
  \centering
  \begin{tabular}{ccc}
	  \includegraphics[width=0.22\textwidth]{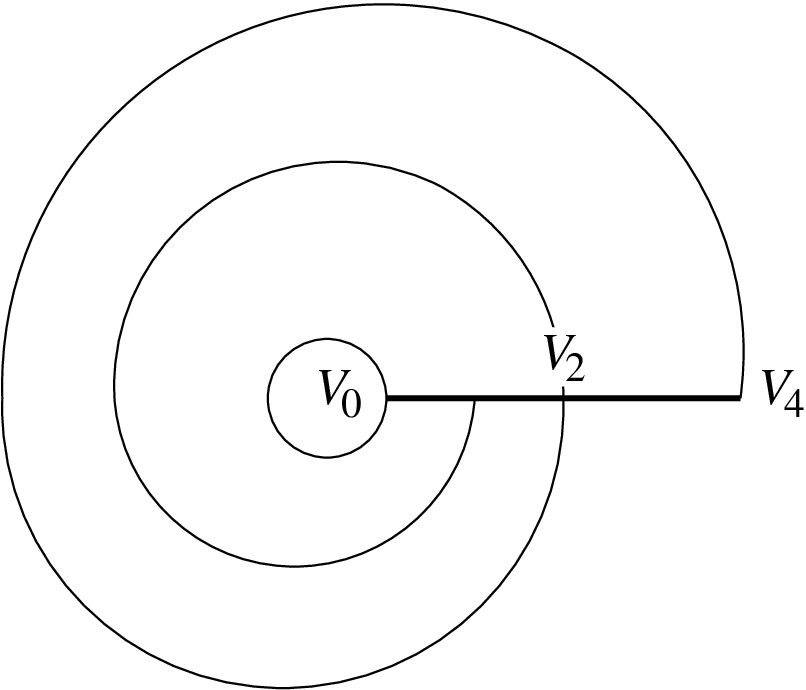} & ~~~ &
	  \includegraphics[width=0.22\textwidth]{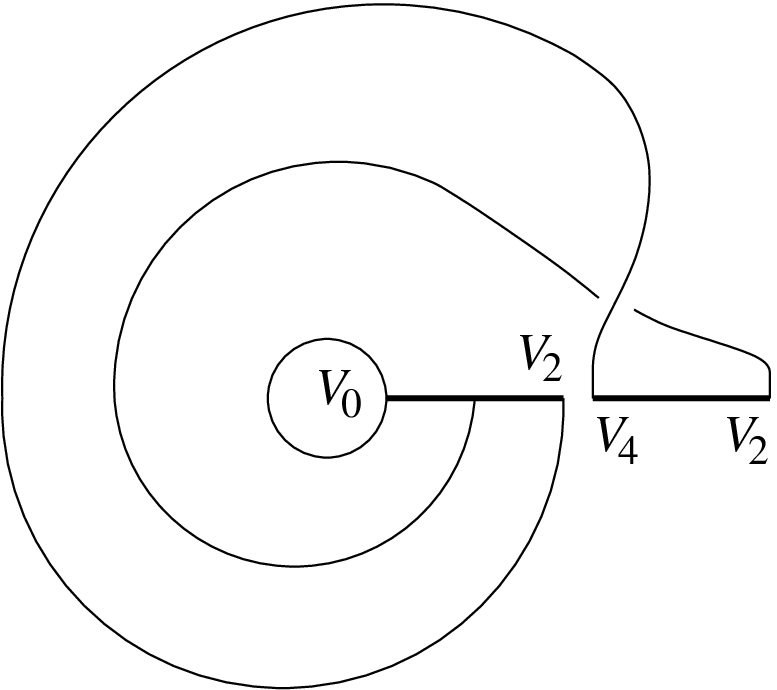} \\[0.1cm]
	  (a) Original strip & &
	  (b) Corrected strip \\[0.2cm]
	  \multicolumn{3}{c}{
             \includegraphics[width=0.33\textwidth]{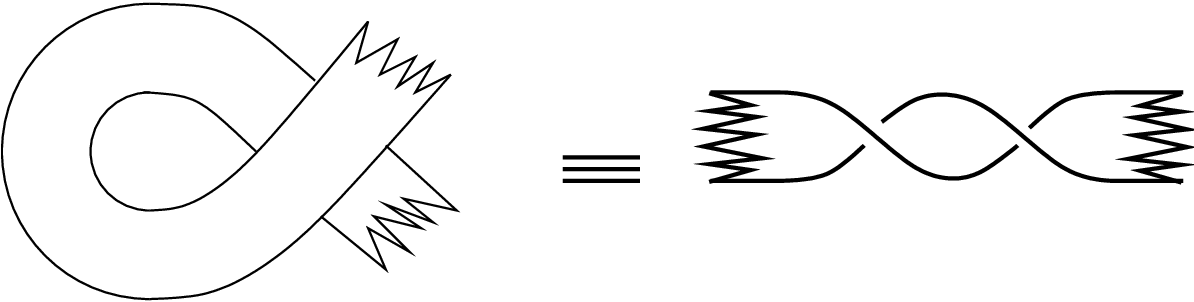}}  \\[0.1cm]
     \multicolumn{3}{c}{  (c) Third Reidemeister move} \\[-0.2cm]
  \end{tabular}
  \caption{Strip $\1$ as originally seen (a) and as corrected when the right 
component of the Poincaré section is correctly oriented. The third 
Reidemeister move explains the origin of the $\pi$-twist.}
  \label{subbrama}
\end{figure}

The template for the Lorenz attractor [Fig.\ \ref{bramalor}(a)] is described by
the linking matrix
\begin{equation}
  L_{ij} = 
  \left[
    \begin{array}{cccc}
      0 & 0 & 0 & 0 \\
      0 & +1 & 0 & 0 \\
      0 & 0 & 0 & 0 \\
      0 & 0 & 0 & +1 
    \end{array}
  \right\rsem
\end{equation}
and the joining matrix
\begin{equation}
  J_{ij} = 
    \begin{array}{l}
      \left[
        \begin{array}{cccc}
		1 & \cdot & \cdot & 1 
        \end{array}
	    \right] \\[0.1cm]
      \left[
        \begin{array}{cccc}
		\cdot & 1 & 1 & \cdot  
        \end{array}
      \right] \, .
    \end{array}
\end{equation}

\noindent
{\underline{Templex}}
\vspace{0.3 cm}

The construction of a Lorenz templex was described in detail in section \ref{sectemplex}: the generating complex and its digraph in $T_2({\rm L})$ are sketched in Figs\ \ref{complorun12}(c-d). If we analyze $K_2({\rm L})$ as a standard complex, one finds a single 0-generator, ${\cal H}_0 = \left[< \displaystyle 0> \right]$, which speaks of a single connected component. There are two 1-generators: 

\[
  \begin{array}{l}
	  {\cal H}_1(K_2({\rm L}))= [[ \displaystyle \langle 4,14 \rangle - \langle 4,20 \rangle +\langle 14,17 \rangle + \langle 17,20 \rangle , \\ \hspace{1.1cm}
    \langle 0,5 \rangle - \langle 0,11 \rangle +\langle 5,8 \rangle + \langle 8,11 \rangle ]]
  \end{array}
\]

\noindent
which identify and locate the two focus-type holes in the attractor. There are no 2-generators (no enclosed cavities): ${\cal H}_2 = \emptyset$. 

What does the generating templex add to this description? It unveils a joining 
locus made of two-components, indicating the existence of a bipartite
Poincaré section. This yields three generatexes and four stripexes $\mathcal S_i$ ($i=1,...4$) which are equivalent to the four strips in the template, two of them have local 
twists (${\cal S}_3$ and ${\cal S}_4$). The three generatexes can be easily 
recognised in the decomposition of the template shown in Figs\ 
\ref{bramalor}(b) and \ref{bramalor}(c): in the latter, the stripexes 
${\cal S}_3$ and ${\cal S}_4$ are weak since only their union defines a 
generatex.

\subsection{The R\"ossler attractor}
\label{rostemp}

\noindent
{\underline{Template}}
\vspace{0.3 cm}

\begin{figure}[ht]
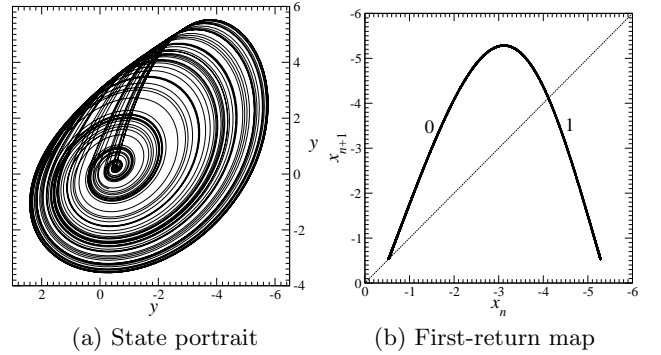

  \centering
  \begin{tabular}{cc}
    \includegraphics[width=0.2280\textwidth]{rosatco.eps} &
    \includegraphics[width=0.23\textwidth]{rosmapco.eps} \\
    (a) State portrait & (b) First-return map \\[-0.2cm]
  \end{tabular}
  \caption{Chaotic attractor produced by the R\"ossler system (\ref{roseq76})
when the symbolic dynamics is nearly complete. Parameter values: $a=0.43295$,
$b = 2$, and $c=4$.} 
  \label{rosko76}
\end{figure}

The topologically simplest chaotic attractor (Fig.\ \ref{rosko76}) is produced 
by the R\"ossler system\cite{Ros76c}
\begin{equation}
  \label{roseq76}
  \left\{
    \begin{array}{l}
      \dot{x} = -y -z \\[0.1cm]
      \dot{y} = x + ay \\[0.1cm]
	    \dot{z} = b + z (x-c)  \, .
    \end{array}
  \right.
\end{equation}
It is characterized by a smooth first-return map to a Poincaré section [Fig.\ \ref{rosko76}(b)].\cite{Let95a} Since the map is smooth and unimodal, the typical route to this chaotic attractor is a period-doubling cascade. This attractor was labelled as C$^1$T$^1$ according to the nomenclature introduced in the taxonomy of chaos.\cite{Let21f} It is bounded by a trivial torus (1,1). 

\begin{figure}[ht]
  \centering
  \begin{tabular}{ccc}
          \includegraphics[height=0.23\textwidth]{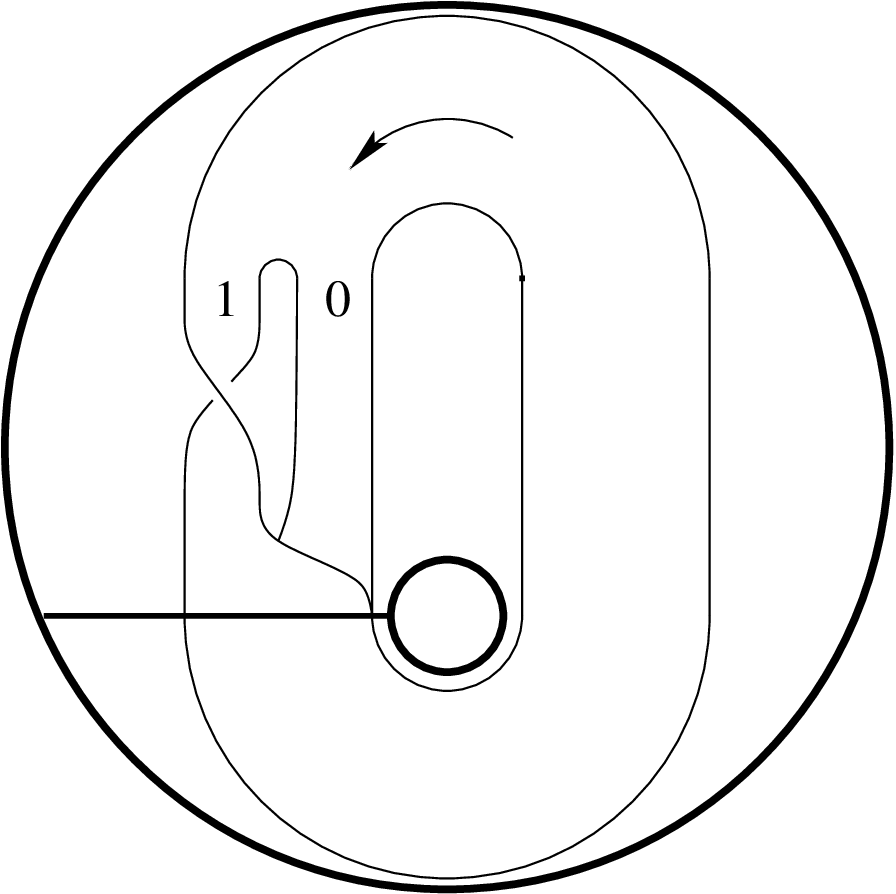} & 
 	  \includegraphics[height=0.25\textwidth]{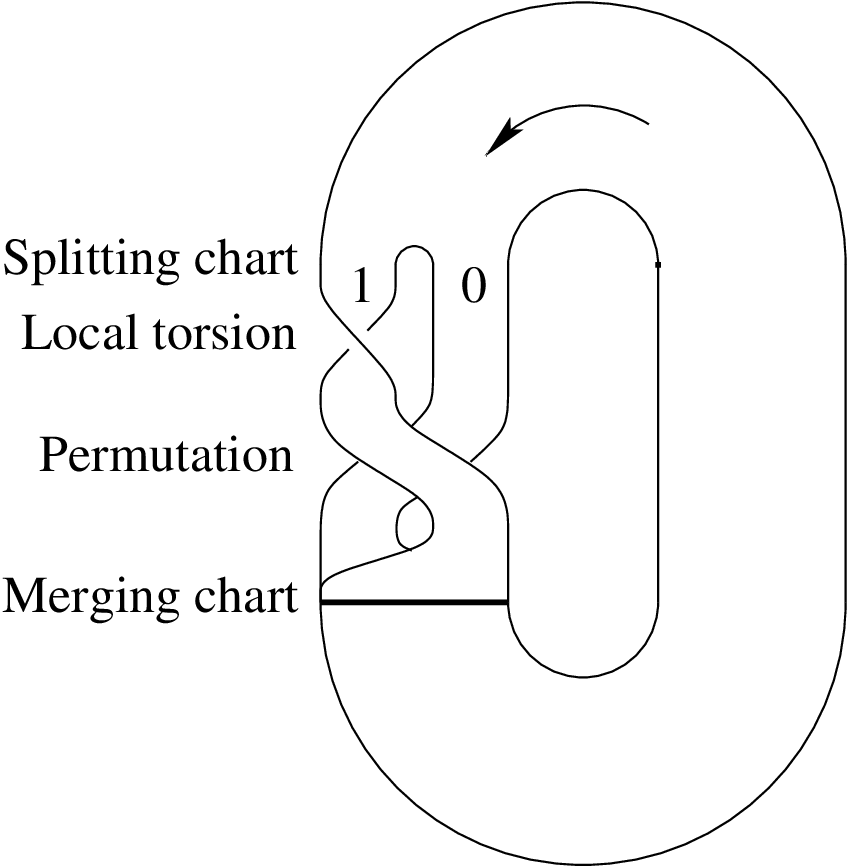}  \\[0.1cm]
	  (a) Direct template & (b) Standard template \\[0.2cm]
	  \includegraphics[width=0.15\textwidth]{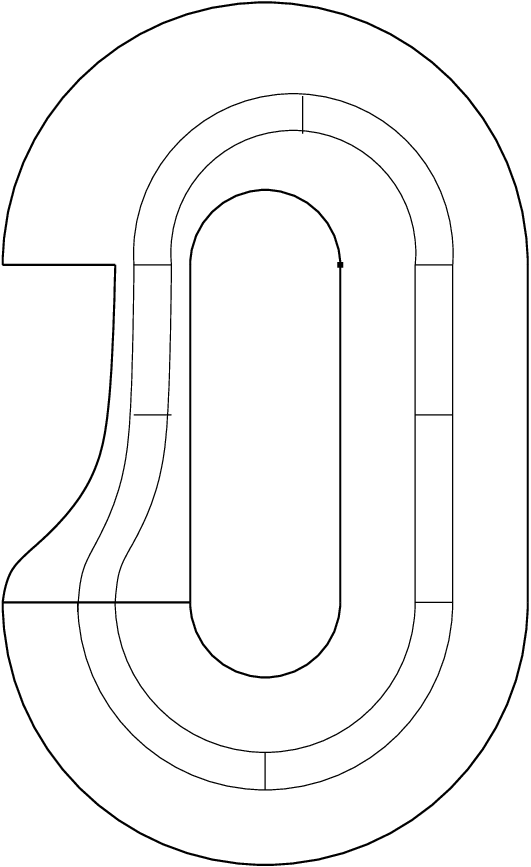} & 
	  \includegraphics[width=0.15\textwidth]{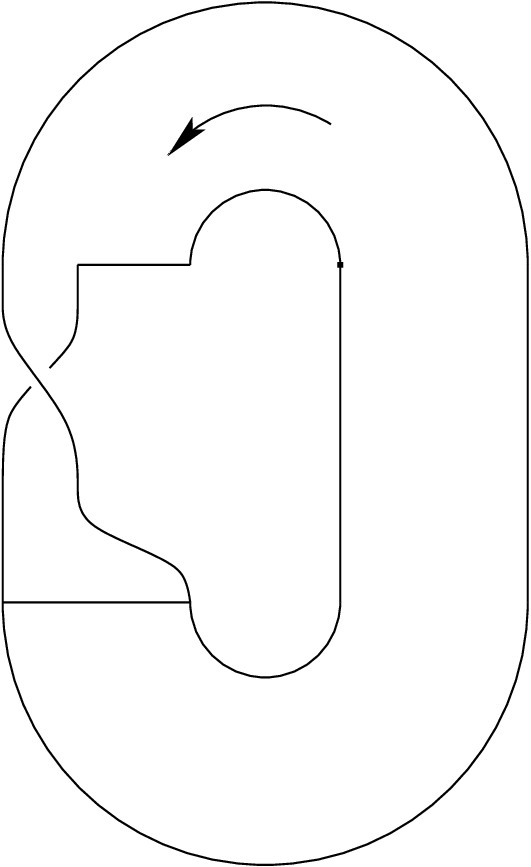} \\[0.1cm]
	  (c) Strip 0 & (d) Strip 1 \\[-0.2cm]
  \end{tabular}
  \caption{Direct (a) and standard (b) template for the R\"ossler attractor plotted in Fig.\ \ref{rosko76}.  The genus-1 bounding torus is drawn in (a) as thick circles. An example of 2-cycle is drawn in the strip 0 (c).}
  \label{bramaros}
\end{figure}

The R\"ossler attractor is described by the template drawn in Fig.\ \ref{bramaros}(c). It should be decomposed into two strips, namely a normal band [Fig.\ \ref{bramaros}(a)] and a M\"obius band [Fig. \ref{bramaros}(b)].\cite{Ros76a} A direct template can be drawn directly from the $y$-$x$ plane projection [note that the $y$-axis in Fig. \ref{rosko76}(a) is inverted for a better equivalence with the template drawn in Fig.\ \ref{bramaros}(a)]. The strips are encoded with the same integers as the branches of the first-return map: an even (odd) integer is used for a strip with an even (odd) local torsion. The natural order $0 < 1$ is drawn from the center to the periphery of the attractor [Fig.\ \ref{bramaros}(b)]. The standard template is described by the linking matrix
\begin{equation}
  L_{ij} = 
  \left[
    \begin{array}{cc}
       0 & -1 \\[0.1cm]
      -1 & -1 
    \end{array}
  \right\rsem \, . 
\end{equation}

\noindent
{\underline{Templex}}
\vspace{0.5 cm}

\begin{figure}[ht]
  \centering
  \begin{tabular}{c}
    \includegraphics[width=0.32\textwidth]{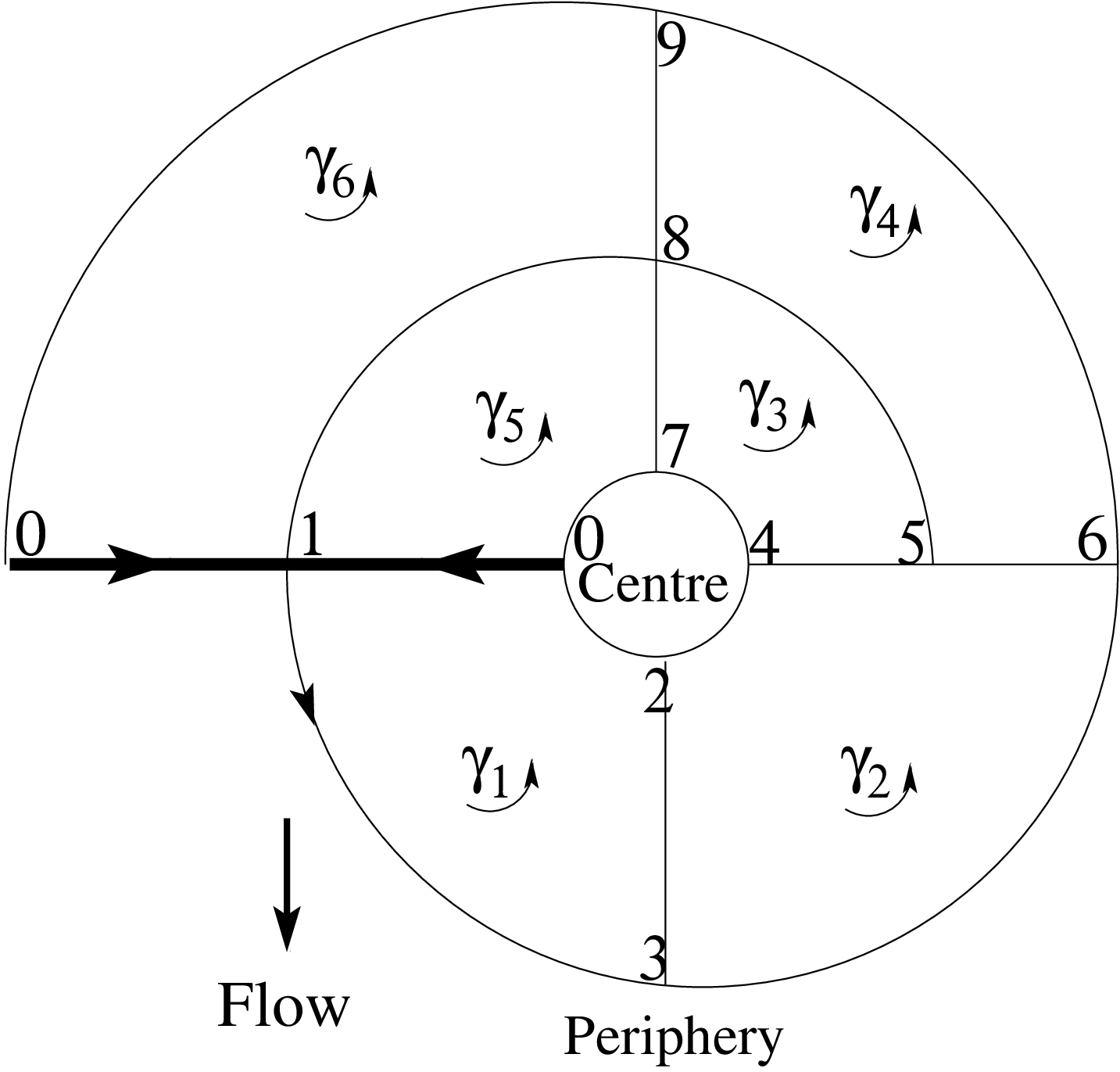} \\
    (a) Generating complex \\[0.2cm]
    \includegraphics[width=0.22\textwidth]{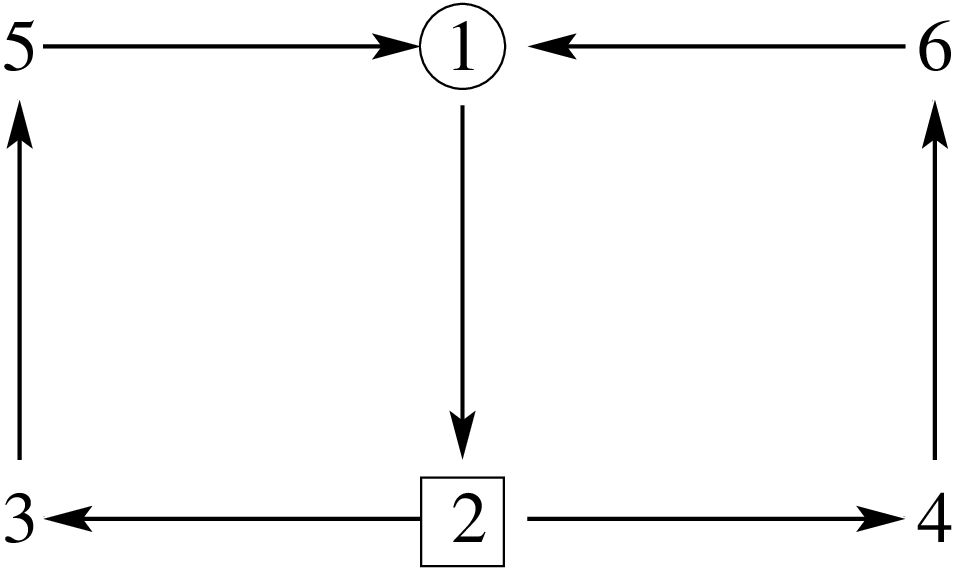} \\
    (b) Generating digraph \\[-0.2cm]
   \end{tabular}
  \caption{Generating templex $T({\rm R})$ for the R\"ossler attractor made of the generating complex $K({\rm R})$ shown as a planar diagram in (a) and the generating digraph $G({\rm R})$ (b). The thick line indicates the joining 1-cell $\langle 0,1 \rangle$. }
  \label{rossler_complex}
\end{figure}

Let us now describe this attractor using a generating templex. As described in section \ref{sectemplex}, the first step is to construct a BraMAH complex. This is performed algorithmically, handling a set of points in the state space. The complex is next endowed with a digraph which reproduces the structure of the flow along the branched 2-manifold, till a generating templex is obtained. The result of the whole procedure is sketched in Fig.\ \ref{rossler_complex} where the complex $K({\rm R})$ is drawn unfolded, that is, in a planar diagram, as is customary in homology theory. In planar diagrams, gluing instructions require some cells to be drawn twice with the same labels and direction. This is the case for the joining 1-cell $\langle 0,1 \rangle$ in $K({\rm R})$, which is drawn twice in Fig.\ \ref{rossler_complex}(a). Flow-orienting this complex is achieved by propagating the orientation from $\gamma_1$ according to our convention [it is obvious in the planar representation shown in Fig.\ \ref{rossler_complex}(a)].

The complex $K({\rm R})$ has a single 0-generator, ${\cal H}_0(K({\rm R})) 
= [< \displaystyle 0 >]$, that is, it has a single connected component. The
single 1-generator of $K({\rm R})$ is
\[ {\cal H}_1(K({\rm R})) 
= [[ \displaystyle  \langle 0, 2 \rangle- \langle 0, 7 \rangle +\langle 2, 4 \rangle + \langle 4, 7 \rangle  ]] \]
which identifies the focus-type hole in the attractor. There are no 2-generators (no enclosed cavities): ${\cal H}_2(K({\rm R})) = \emptyset$. Notice that, 
homologically, $K({\rm R})$ is a cylinder: fortunately, the generating templex 
tells us much more about its structure. 

The generating templex $T({\rm R})= (K({\rm R}),G({\rm R}))$ has the following 
properties. There is one joining 2-cell ($\gamma_1$). Since there is a single 
joining 1-cell, the joining locus has a single component (as expected for an 
attractor bounded by a genus-1 torus) and there is no splitting 0-cell. There 
are two stripexes $\mathcal{S}_1$ and ${\cal S}_2$ associated with the cycles
\[
  \begin{array}{l}
     c_1 \equiv \underline{1} \rightarrow 2 \rightarrow 4 \rightarrow 6  
	  \rightarrow \underline{1} \\
     c_2 \equiv \underline{1} \rightarrow 2 \rightarrow 3 \rightarrow 5  
	  \rightarrow \underline{1} \\
  \end{array}
\]
The associated generatexes are simple ($p = 1$) and correspond to strips 0 and 
1, respectively. The stripex ${\cal S}_1$ is a M\"obius band -- it has some
torsion elements ($b_1=2$ for $\sigma_1^1= \langle 0,1 \rangle$ in 
$\partial \Gamma$) --- while $K_2$ is a cylinder or normal band (without 
torsion). 

\begin{figure}[ht]
  \centering
  \includegraphics[width=0.26\textwidth]{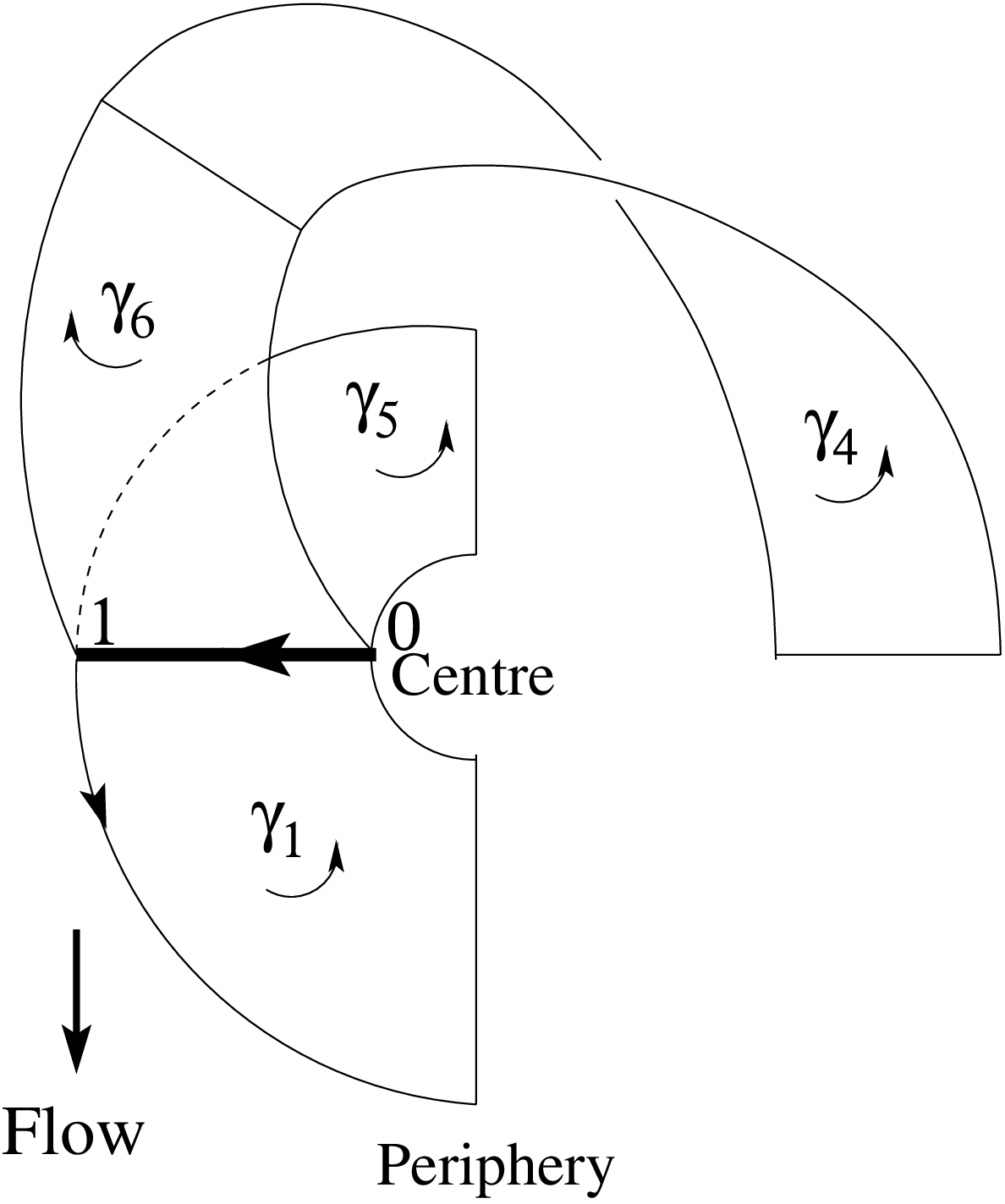} \\[-0.3cm]
  \caption{The three 2-cells ($\gamma_1$, $\gamma_5$, and $\gamma_6$) sharing the joining 1-cell. The 2-cell $\gamma_4$ is added to show the local twist of 
the stripex $\mathcal{S}_1$.}
  \label{rossler_2chain}
\end{figure}

The stripex ${\cal S}_1$ has a local twist which is sketched in Fig.\ 
\ref{rossler_2chain}. According to our convention for the top, the 2-cell $\gamma_6$ is at the top of the cell $\gamma_5$. 

\subsection{The three-strips Rossler attractor}

\noindent
{\underline{Template}}
\vspace{0.3 cm}

\begin{figure}[ht]
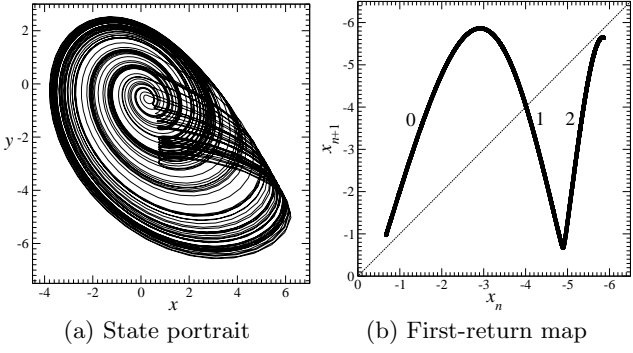

  \centering
  \begin{tabular}{cc}
    \includegraphics[width=0.2280\textwidth]{rosatt3b.eps} &
    \includegraphics[width=0.23\textwidth]{rosmap3.eps} \\
    (a) State portrait & (b) First-return map \\[-0.2cm]
  \end{tabular}
  \caption{Chaotic attractor with three branches as produced by the R\"ossler 
system (\ref{roseq76}). Parameter values: $a=0.492$, and other parameter values
	as in Fig.\ \ref{rosko76}.}
  \label{rosko3b}
\end{figure}

When the parameter $a$ of the R\"ossler system is set to 0.492, the 
first-return map to a Poincaré section has three branches (Fig.\ 
\ref{rosko3b})\cite{Let95a} and the template is drawn in Fig.\ \ref{extemp}. 
This attractor is labeled R$_3$. 

\noindent
{\underline{Templex}}
\vspace{0.3 cm}

A classical homological analysis of $K({\rm R}_3)$ yields the following results: there is, as usual, one 0-generator (one connected component). The 1-generator (corresponding to the hole of the focus type) is given by: ${\cal H}_1(K({\rm R}_3)) = [[ \displaystyle  \langle 0, 2 \rangle- \langle 0, 6 \rangle +\langle 2, 4 \rangle + \langle 4, 6 \rangle ]]$ which identifies the focus-type hole in the attractor.
No 2-generators are found (there are no cavities enclosed).

\begin{figure}[ht]
  \centering
  \includegraphics[width=0.40\textwidth]{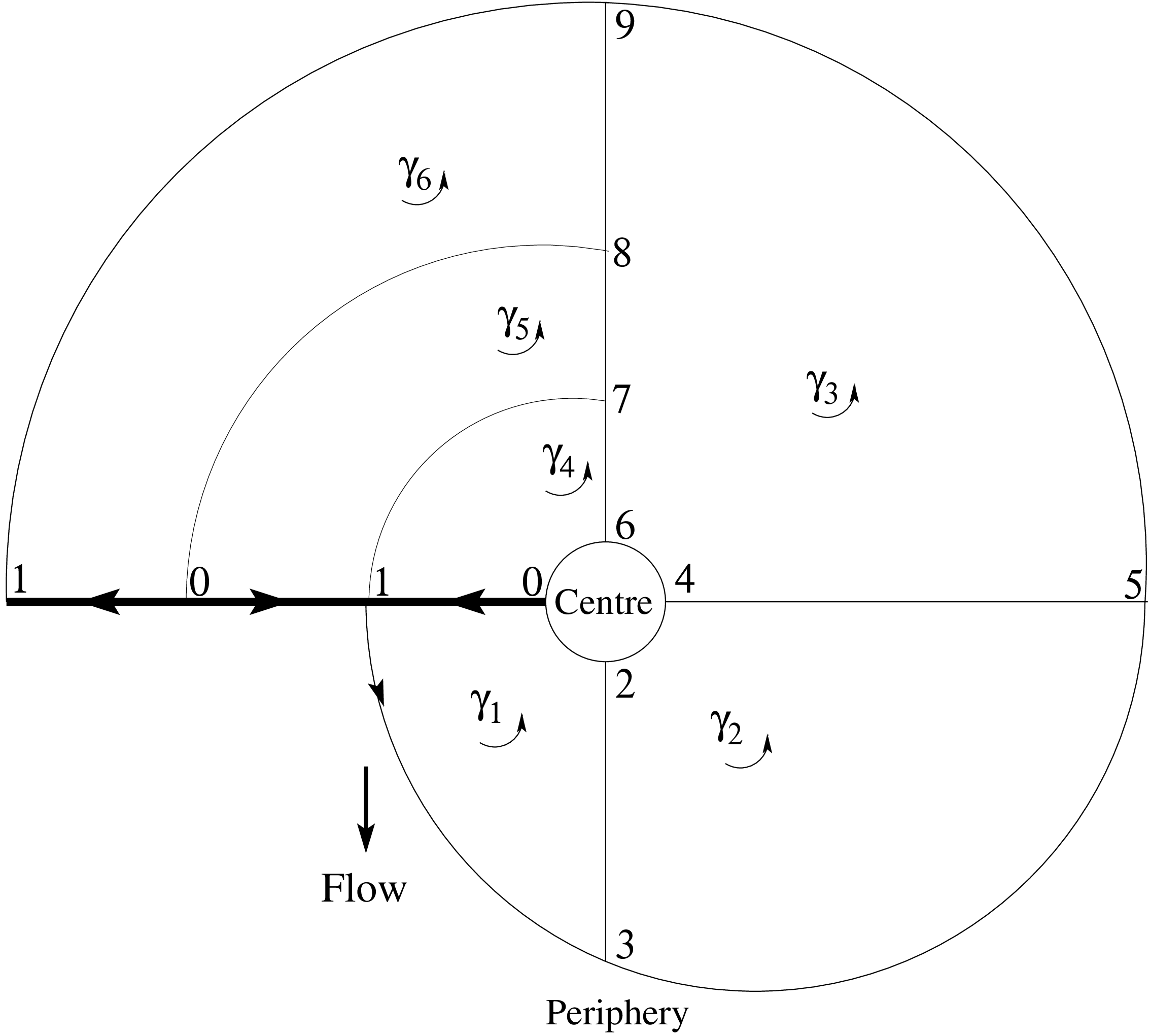}  \\
	(a) Generating complex $K({\rm R}_3)$\\[0.2cm]
  \includegraphics[width=0.14\textwidth]{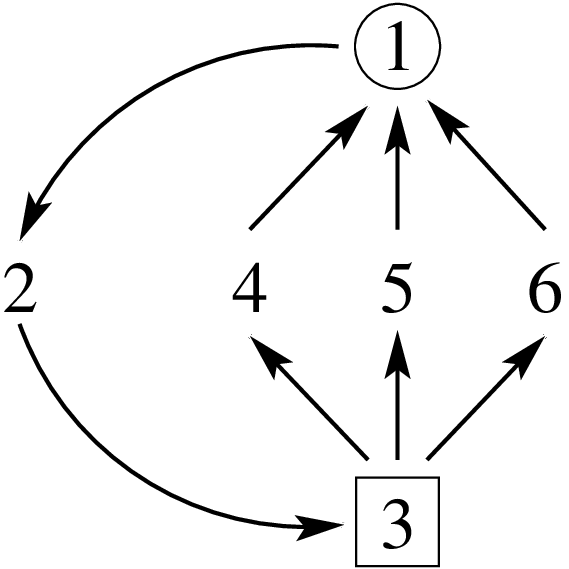}  \\
	(b) Generating digraph $G({\rm R}_3)$ \\[-0.2cm]
  \caption{Generating templex $T({\rm R}_3)=(K({\rm R}_3),G({\rm R}_3))$ for the three-strip Rossler attractor. Parameter values: $a = 0.492$, $b = 2$ and $c = 4$.}
  \label{rossler3_complex}
\end{figure}

A templex for the three-strips R\"ossler attractor 
$T({\rm R}_3)=(K({\rm R}_3),G({\rm R}_3))$ is proposed in Fig. 
\ref{rossler3_complex}. The joining locus has one component and it is made of 
the 1-cell $\langle 0,1 \rangle$. The 2-cell $\gamma_1$ is the unique outgoing 
joining 2-cell, and $\gamma_4$, $\gamma_5 $, and $\gamma_6$ are the three
ingoing joining 2-cells. 

As expected, three generatexes and three stripexes $\mathcal{S}_{1,2,3}$ are found ($N_{\rm s} = 3$). They are associated
with the three cycles 
\[
  \begin{array}{l}
    c_1 \equiv \underline{1} \rightarrow 2 \rightarrow 3 \rightarrow 4  
	  \rightarrow \underline{1} \\
    c_2 \equiv \underline{1} \rightarrow 2 \rightarrow 3 \rightarrow 5  
	  \rightarrow \underline{1} \\
    c_3 \equiv \underline{1} \rightarrow 2 \rightarrow 3 \rightarrow 6  
	  \rightarrow \underline{1} 
  \end{array}
\]
which are indeed non-equivalent because they are associated with different
ingoing nodes. Of the the three stripexes, only $\mathcal{S}_2$ has a local twist, corresponding to the intermediate strip of the template in Figure 3. $\mathcal{S}_1$ corresponds to the strip with no local torsion and $\mathcal{S}_3$ to the strip with an even parity.

\subsection{The Burke and Shaw attractor}

\noindent
{\underline{Template}}
\vspace{0.3 cm}

\begin{figure}[ht]
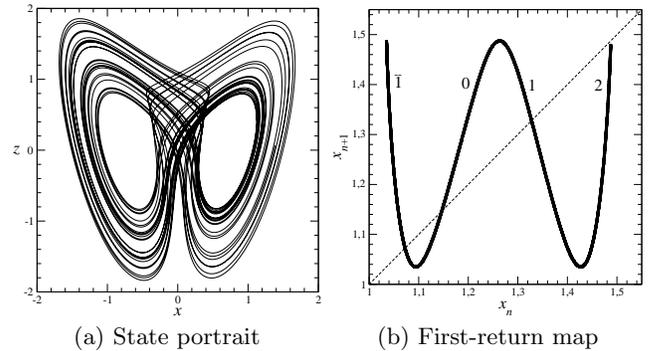

  \centering
  \begin{tabular}{cc}
    \includegraphics[width=0.23\textwidth]{shawatt.eps} &
    \includegraphics[width=0.23\textwidth]{shamap.eps} \\
    (a) State portrait & (b) First-return map \\[-0.2cm]
  \end{tabular}
  \caption{Chaotic attractor produced by the Burke and Shaw system (\ref{Shaweq81}) when the symbolic dynamics is nearly complete. Parameter values: $a=10$, and $b = 2.271$.}
  \label{burkshaw}
\end{figure}

\begin{figure}[ht]
  \centering
  \begin{tabular}{cc}
    \begin{tabular}{c}
      \includegraphics[width=0.24\textwidth]{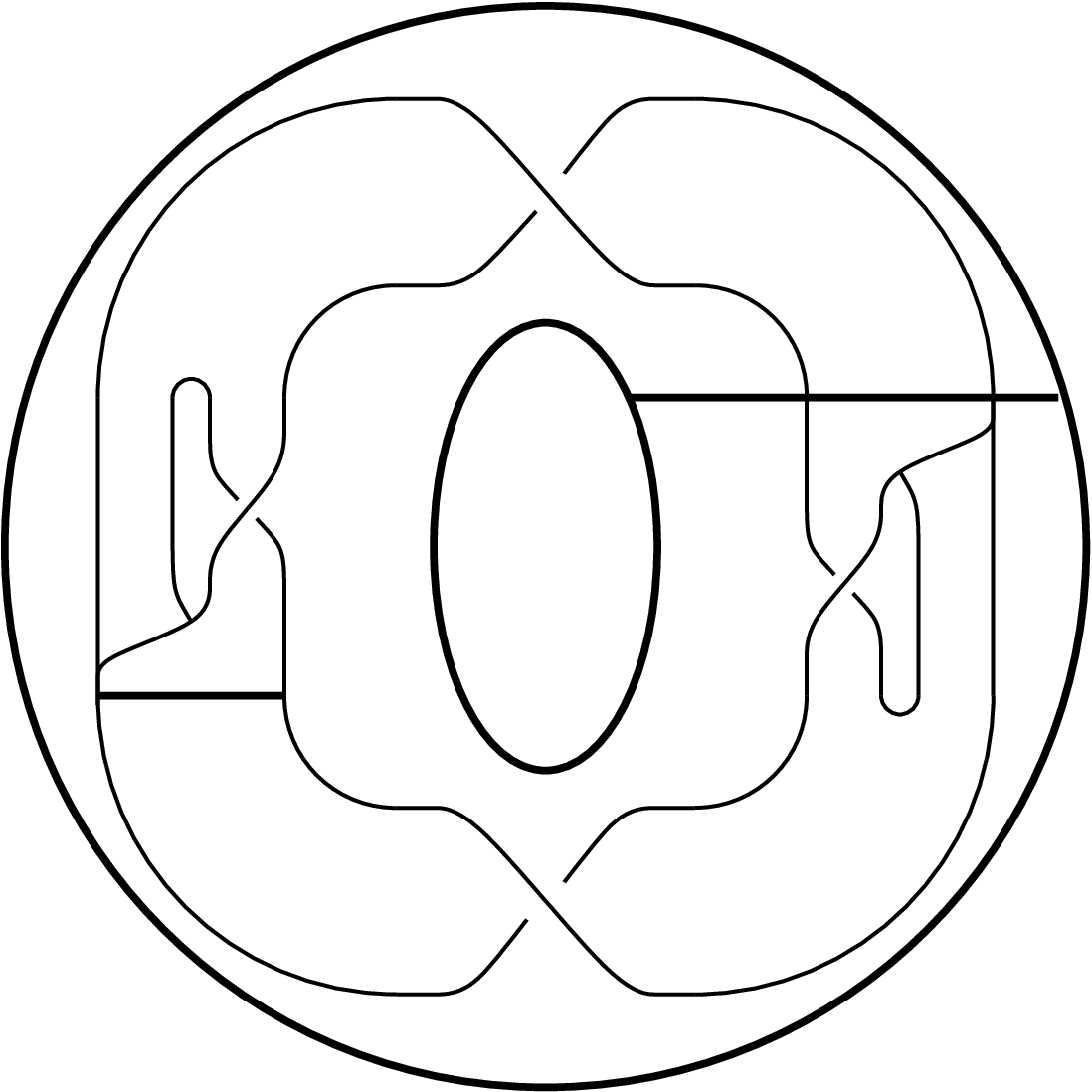} \\
      (a) Direct double template \\[0.2cm]
      \includegraphics[width=0.18\textwidth]{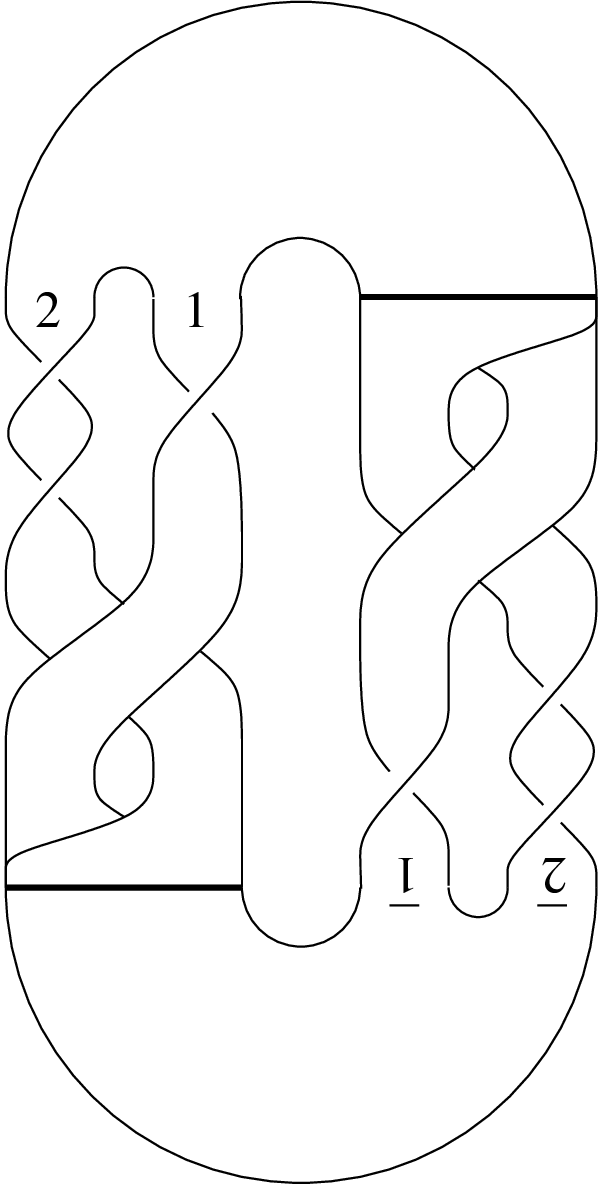} \\
      (b) Reduced double template 
    \end{tabular} & \\[-10.0cm]
    & \includegraphics[width=0.16\textwidth]{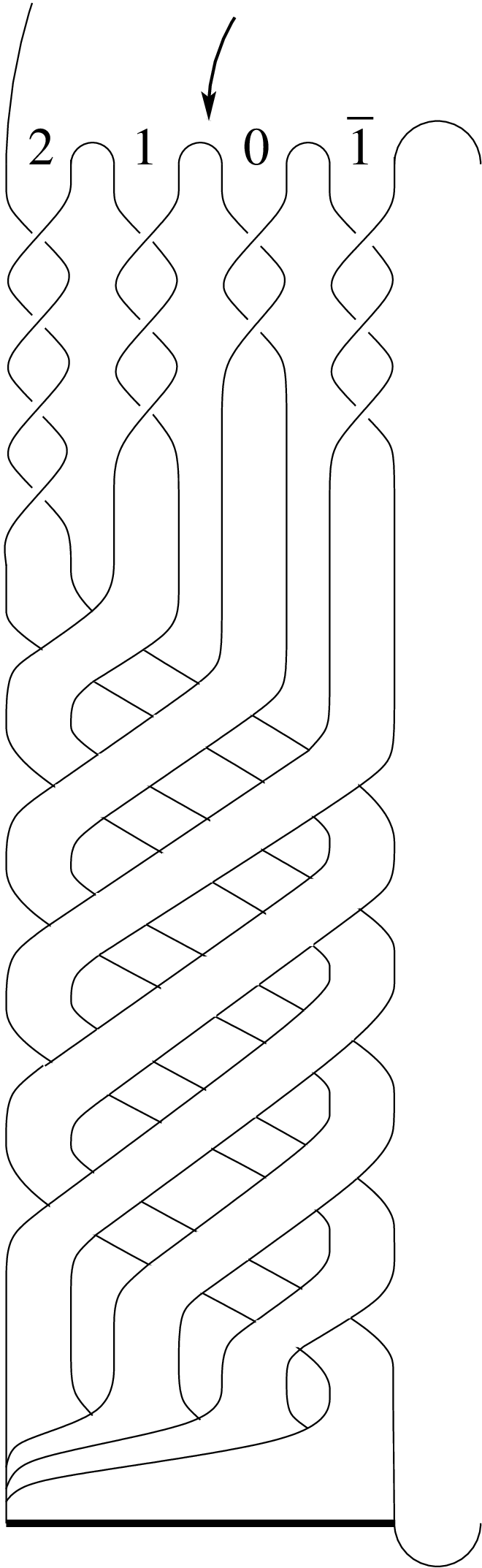} \\[0.08cm]
    & (c) Template \\[-0.1cm]
  \end{tabular}
  \caption{Double template and genus-1 bounding torus (a) for the Burke \& Shaw attractor plotted in Fig.\ \ref{burkshaw}(a). Direct double template (a), 
	reduced double template (b) and reduced template (c).}
  \label{bramaBS}
\end{figure}

There is a Lorenz-like system which produces a quite common attractor: 
this is the Burke and Shaw 
system \cite{Sha81,Let96a}
\begin{equation}
  \label{Shaweq81}
  \left\{
    \begin{array}{l}
      \dot{x} = - a x - a y \\
      \dot{y} = - y - a xz \\
      \dot{z} = b + a xy \, . 
    \end{array}
  \right.
\end{equation}
As the Lorenz system (\ref{loreq63}), this system is equivariant under a 
rotation symmetry ${\cal R}_z (\pi)$. Close to the original parameter values,\cite{Sha81} there is a specific attractor which is encountered in many Lorenz-like systems.\cite{Let05c} Indeed, for $a = 10$ and $b = 2.271$, the attractor plotted in Fig.\ \ref{burkshaw}(a) is characterized by a four-branches first-return map [Fig.\ \ref{burkshaw}(b)]. It has four branches and a possible natural order between the integers used for labelling them is $\1 < 0 < 1 < 2$.\cite{Let96a} This attractor is bounded by a genus-1 torus as sketched around the template drawn in Fig.\ \ref{bramaBS}(a). As a consequence, the Poincaré section should have a single component. Nevertheless, as shown in Fig.\ \ref{burkshaw}(a), there are two foldings located in the neighborhood of the rotation axis, leading to two joining charts.

It is actually possible to sketch the topology of the attractor as the product 
of two mixers [each of them being located between one splitting chart and one joining chart as drawn in Figs. \ref{bramaBS}(a) and \ref{bramaBS}(b)]. As exhibited in the Lorenz attractor with the third Reidemeister move [Fig. \ref{subbrama}(c)], there is a global torsion (a global torsion differs from a local one in the sense that it is applied to all the strips of the attractor, the latter being applied to a single strip, by definition) between a sequence of one joining chart followed by one splitting chart. With the help of the Reidemeister moves, each global torsion can be moved in one of the two mixers, to get the reduced double template shown in Fig.\ \ref{bramaBS}(b). In this  representation, the rotation symmetry is obvious.

It is possible to conjugate these two mixers into a single one. In this case, the two strips of the first (say the left) mixer are spread into the two strips of the second (right) mixer to form four strips. Propagating permutations and local torsions using Reidemeister moves leads to the reduced template drawn in Fig.\ \ref{bramaBS}(c): it is described by the linking matrix
\begin{equation}
  L_{ij} = 
  \left|
    \begin{array}{cccc}
       3 &  2 &  2 &  3 \\[0.1cm]
       2 &  2 &  2 &  3 \\[0.1cm]
       2 &  2 &  3 &  3 \\[0.1cm]
       3 &  3 &  3 &  4
     \end{array}
   \right\rsem
\end{equation}

\begin{figure}[ht]
  \centering
  \includegraphics[width=0.30\textwidth]{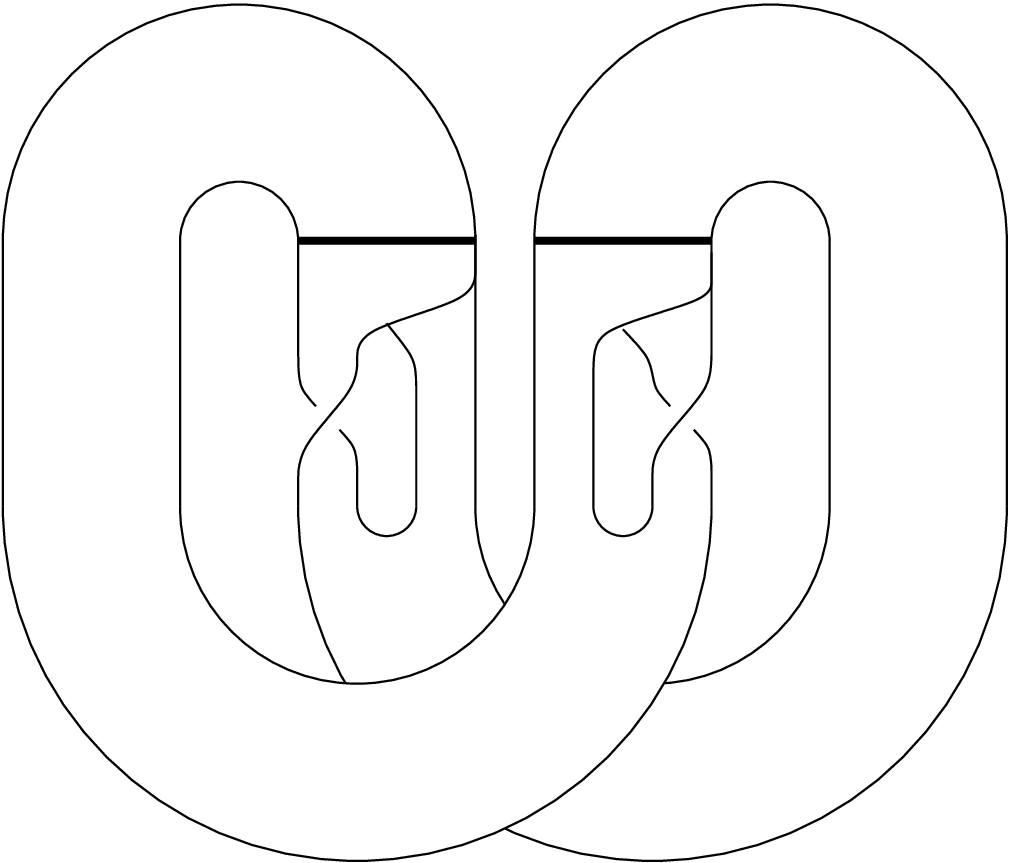} \\
  (a) Direct template \\[0.2cm]
  \caption{Double templates for the Burke and Shaw attractor.}
  \label{templadirBS}
\end{figure}

For a more direct correspondence with templexes, it is useful to redraw the direct template as shown in Fig.\ \ref{templadirBS}: it is topologically equivalent to the
template drawn in Fig.\ \ref{bramaBS}.

\vspace{0.3 cm}
\noindent
{\underline{Templex}}
\vspace{0.3 cm}

\begin{figure}[ht]
  \centering
  \includegraphics[width=0.49\textwidth]{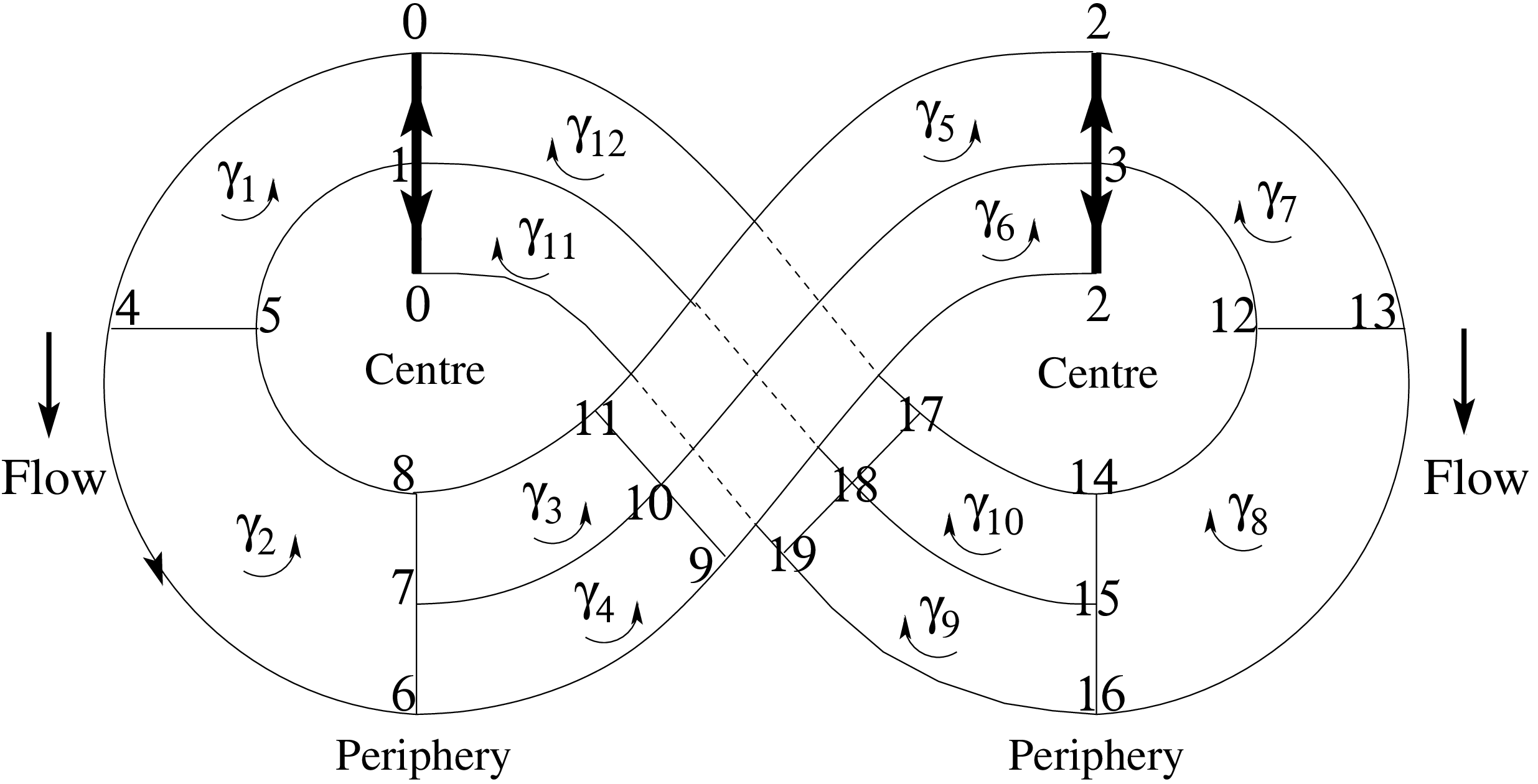} \\
	(a) Generating complex \\[0.2cm]
  \includegraphics[width=0.18\textwidth]{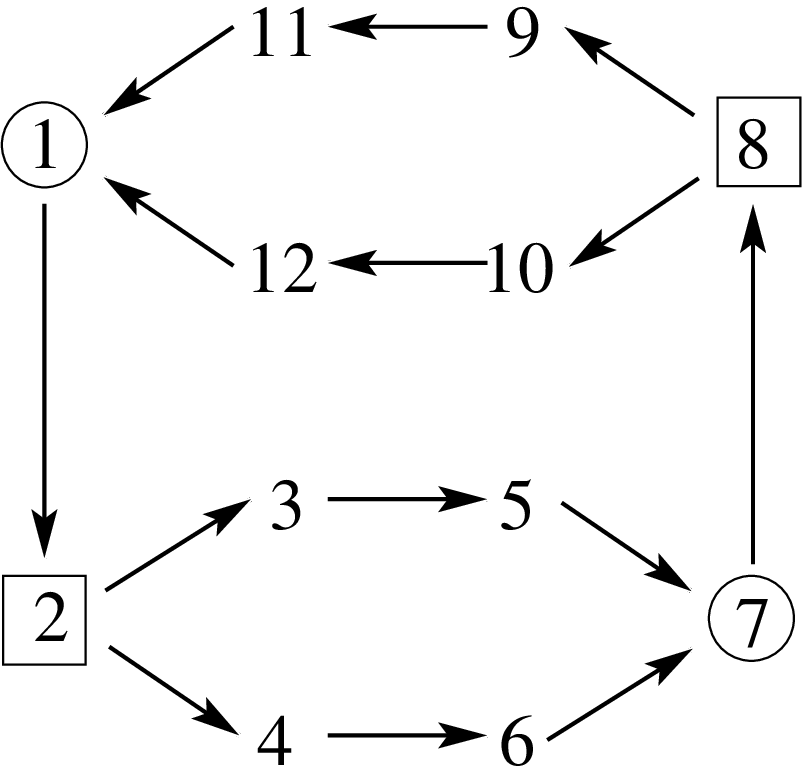} \\
	(b) Generating digraph \\[-0.2cm]
  \caption{Templex for the Burke and Shaw attractor: generating complex (a) 
endowed with a directed graph (b). There are two joining 1-cells (thick lines 
with arrows). They are no splitting 0-cells.}
  \label{BS_complex}
\end{figure}

Let us consider the complex $K({\rm BS})$ shown in Fig.\ \ref{BS_complex}(a). 
Computing the homology groups, one finds a single 0-generator (one connected component), and a single 1-generator (corresponding to the focus-type hole in the attractor). 
\[
  \begin{array}{l}
    {\cal H}_1(K({\rm BS})) = [[ \displaystyle \langle 0,4\rangle - \langle 0,17 \rangle +\langle 2,3 \rangle - \langle 2,9 \rangle + \\ \hspace{1.1cm}
    + \langle 3,12 \rangle + \langle 4,6 \rangle +\langle 6,9 \rangle + \langle 12,14 \rangle  + \langle 14,17 \rangle ]]
  \end{array}
\]
As expected, no cavities are found. 

The generating complex and digraph forming templex $T({\rm BS})=(K({\rm BS}),G({\rm BS}))$ for this attractor are presented in Fig.\ \ref{BS_complex}. The orientation of the generating 
complex $K({\rm BS})$ is started from the two joining 2-cells: $\gamma_1$ and 
$\gamma_8$. The analysis of the templex yields a joining locus with the two 
components $J_1({\rm BS})=\langle 1,0 \rangle$ and 
$J_2({\rm BS})=\langle 3,2 \rangle$, which correspond to the two components of the Poincaré section in the direct template [Fig.\ \ref{bramaBS}(b) or
Fig.\ \ref{templadirBS}]. There are four stripexes $\mathcal{S}_{1,2,3,4}$, corresponding to the following weak cycles 
\[
  \begin{array}{l}
    c_{1_1} \equiv \underline{1} \rightarrow 2 \rightarrow 3 \rightarrow 5 
	  \rightarrow \underline{7} \\
    c_{1_2} \equiv 
    \underline{7} \rightarrow 8 \rightarrow 9 \rightarrow 11  
	  \rightarrow \underline{1} \\
    c_{2_1} \equiv 
      \underline{1} \rightarrow 2 \rightarrow 4 \rightarrow 6 
	  \rightarrow \underline{7} \\
    c_{2_2} \equiv 
      \underline{7} \rightarrow 8 \rightarrow 10 \rightarrow 12 
	  \rightarrow \underline{1} \\
  \end{array}
\]
Each pair of weak cycles comes from a degenerated generatex, of order 2.  $\mathcal{S}_{1,4}$ have local twists, while $\mathcal{S}_{2,3}$ do not. In the semi-planar diagram, $\gamma_{5,6}$ are on the top of $\gamma_{11,12}$.

\section{Conclusion}
\label{conc}

A new concept, named templex, is here introduced with the scope of extending 
the description of chaotic attractors in terms of cell complexes and homology 
groups. The two missing elements in the  description in terms of a complex are the computation of the joining locus at which more than two $\kappa$-cells can be attached, and the direction of the flow, now used to orient the cells of the complex in a particular manner. The underlying flow implies 
that there are only some transitions between cells which are possible. This 
information is encoded with a directed graph whose nodes are the highest 
dimensional cells (the $\kappa$-cells) of the complex. The digraph is the key 
companion of the cell complex, both forming a templex. The templex contains all the information that is necessary to decompose the the original complex into  an invariant number of smaller units, the analogs of the strips in a template.

We explained how a templex can be built from a BraMAH complex to obtain a 
generating templex, leading naturally to a decomposition of the templex into 
stripexes, as strips are the units of template. The procedure was successfully 
applied to three inequivalent chaotic attractors, namely the 
Lorenz attractor, the R\"ossler attractor, the three-strip R\"ossler attractor 
and the Burke-and-Shaw attractor. The equivalence between template and 
generating templex is thus illustrated but with a major difference: there is no
intrinsic limitation to branched 2-manifolds in the construction of a templex. 
The gate to a topological characterization of chaotic attractors produced by
dynamical systems whose dimension is greater than three is therefore opened:
this is currently under consideration and will be reported in forthcoming 
papers.


\acknowledgements
Some of the ideas that led to the consideration of the templex came from stimulating discussions with Ivan Sadofschi Costa. We also thank Sylvain Mangiarotti for enriching discussions.  This work is supported by the French National program LEFE (Les Enveloppes Fluides et l’Environnement) and by the CLIMAT-AMSUD 21-CLIMAT-05 project (D.S.). G.D.C. gratefully acknowledges her postdoctoral scholarship from CONICET.

\appendix
\section{Computing the homology groups of the torus}

\begin{figure}[ht]
  \centering
  \includegraphics[width=0.22\textwidth]{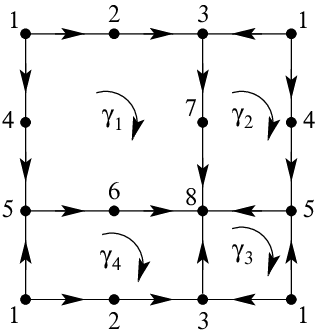} \\[-0.2cm]
  \caption{Complex $K(T^2)$ for the torus. The 0-cells are arbitrarily numbered. The 1-cells are oriented according to this numbering (see the arrows) and the four 2-cells are clockwise.} 
  \label{torusK2}
\end{figure}
Let $K(T^2)$ denote a complex whose underlying space is the torus T$^2$ (Fig.\ \ref{torusK2}). Let all the 2-cells 
$\gamma_i$ 
($i=1,2,3,4$) be clockwise oriented. 
Every 0-cell is a 0-cycle since, by definition, $\partial_0(C_0) =0$. All the 0-cells are homologous because one can reach all of them travelling through the 1-cells of the complex, and there is a single 0-generator.  The four 2-cells are bounded by the 1-chains
\[ 
  \left\{ 
    \begin{array}{l}
      \displaystyle \partial_2{(\gamma_1)}={C}_1^1 =\langle 1,2 \rangle + \langle 2,3 \rangle+ \langle 3,7 \rangle + \langle 7,8 \rangle-\langle 6,8 \rangle - \langle 5,6\rangle\\
      -\langle 4,5 \rangle- \langle 1,4 \rangle\\[0.1cm]
      \partial_2{(\gamma_2)}=\displaystyle {C}_1^2 = -\langle 1,3 \rangle + \langle 1,4 \rangle+ \langle 4,5 \rangle + \langle 5,8 \rangle-\langle 7,8 \rangle - \langle 3,7 \rangle\\[0.1cm]
      \partial_2{(\gamma_3)}=\displaystyle {C}_1^3 =- \langle 5,8 \rangle - \langle 1,5 \rangle+\langle 1,3\rangle +\langle 3,8 \rangle\\[0.1cm]
      \partial_2{(\gamma_4)}=\displaystyle { C}_1^4 = \langle 5,6 \rangle + \langle 6,8 \rangle- \langle 3,8 \rangle -\langle 2,3 \rangle - \langle 1,2 \rangle+ \langle 1,5 \rangle,\\[0.1cm]
    \end{array}
  \right.
  \]
respectively. By definition, these boundaries [$\partial_1(C^i_1)=0$] are 
1-cycles and form the set  ${\cal B}_1 (K(T^2))=\mathrm{im} (\partial_{2})$.
%
In order to compute the set of 1-generators, let us first compute ${\cal Z}_1$, i.e. the kernel of the transpose of $M_1$, in which only non-zero elements are reported:

\begin{equation}
  M_1^T =
  \begin{array}{crrrrrrrr}
    \hline \hline
	  \\[-0.3cm]
    \partial_1 & \langle 1  \rangle&\langle 2  \rangle &\langle 3  \rangle&\langle 4  \rangle&\langle 5 \rangle &\langle 6  \rangle&\langle 7  \rangle& \langle 8 \rangle
	  \\[0.1cm]
    \hline
	  \\[-0.3cm]
	  \langle 1,2 \rangle & -1 & 1 & & & & & &\\
	  \langle 1,3 \rangle & -1 & & 1 & & & & &\\
	  \langle 1,4 \rangle & -1 & & & 1 & & & &\\
	  \langle 1,5 \rangle & -1 & & & & 1 & & &\\
	  \langle 2,3 \rangle & & -1 & 1 & & & & &\\
	  \langle 3,7 \rangle & & & -1 & & & & 1 & \\
	  \langle 3,8 \rangle & & & -1 & & & & & 1\\
	  \langle 4,5 \rangle & & & & -1 & 1 & & &\\
	  \langle 5,6 \rangle & & & & & -1 & 1 & &\\
	  \langle 5,8 \rangle & & & & & -1 & & & 1\\
	  \langle 6,8 \rangle & & & & & & -1 & & 1\\
	  \langle 7,8 \rangle & & & & & & & -1 & 1 \\[0.1cm]
    \hline \hline
  \end{array}
\end{equation}

\begin{equation}
  \left\{
    \begin{array}{l}
     {C}_1^5 =
      \langle 1,2 \rangle + \langle 2,3 \rangle - \langle 1,3 \rangle \\[0.1cm]
      {C}_1^6 =
      -\langle 1,3 \rangle + \langle 1,5 \rangle - \langle 3,8 \rangle +\langle 5,8 \rangle \\[0.1cm]
      { C}_1^7 =
      \langle 1,4 \rangle + \langle 4,5 \rangle - \langle 1,5 \rangle \\[0.1cm]
      {C}_1^8 =
      \langle 3,7 \rangle + \langle 7,8 \rangle - \langle 3,8 \rangle \\[0.1cm]
      {C}_1^9 =
      -\langle 1,3 \rangle + \langle 1,5 \rangle - \langle 3,8 \rangle+\langle 5,6 \rangle+\langle 6,8 \rangle  . 
    \end{array}
  \right.
\end{equation}


\noindent
Let us show that $C_1^5 \sim {C}_1^9-{C}_1^6$. They are homologous if there is a 2-chain ${C}_2$ such that $\partial_2(C_2)=C_1^5 - (C_1^9-{C}_1^6)$. The action of the boundary operator  $\partial_2$ can be written in the matrix form $M_2$. This matrix is used to compute ${\cal B}_1 (K(T^2))$, and therefore ${C}_1^6 \in {\cal B}_1 (K(T^2))$.
\begin{equation}
  M_2 =
  \begin{array}{crrrr}
    \hline \hline
	  \\[-0.3cm]
    \partial_2 & \gamma_1 & \gamma_2 & \gamma_3 & \gamma_4 
	  \\[0.1cm]
    \hline
	  \\[-0.3cm]
	  \langle 1,2 \rangle &  1 &    &    & -1 \\
	  \langle 1,3 \rangle &    & -1 &  1 & \\
	  \langle 1,4 \rangle & -1 &  1 &    & \\
	  \langle 1,5 \rangle &    &    & -1 &  1 \\
	  \langle 2,3 \rangle &  1 &    &    & -1 \\
	  \langle 3,7 \rangle &  1 & -1 &    & \\
	  \langle 3,8 \rangle &   &  &   1 & -1 \\
	  \langle 4,5 \rangle & -1 &  1 &    & \\
	  \langle 5,6 \rangle & -1 &    &    &  1 \\
	  \langle 5,8 \rangle &   &  1 &  -1  &  \\
	  \langle 6,8 \rangle & -1  &    &    & 1\\
	  \langle 7,8 \rangle &  1 & -1 &    & \\[0.1cm]
    \hline \hline
  \end{array}
\end{equation}
in which only non-zero elements are reported. Using the 2-chain $\gamma_1 + \gamma_2$, we get
\begin{equation*}
  \begin{array}{rl}
	  \partial_2(\gamma_1 + \gamma_2) & = \displaystyle
	  \langle 1,2 \rangle
	  + \langle 2,3 \rangle
	  - \langle 1,3 \rangle
	  + \langle 1,4 \rangle
	  + \langle 4,5 \rangle\\
	  & ~~~ + \langle 5,8 \rangle
	  - \langle 6,8 \rangle
	  - \langle 5,6 \rangle
	  - \langle 4,5 \rangle
	  - \langle 1,4 \rangle\\
	  & = \langle 1,2 \rangle
	  + \langle 2,3 \rangle
	  - \langle 1,3 \rangle
	  - \left( \displaystyle \langle 5,6 \rangle
	  + \langle 6,8 \rangle
	  - \langle 5,8 \rangle \right)\\
	  & = {C}_1^5 - ({C}_1^9-{C}_1^6)  
       \end{array}
\end{equation*}
Similarly, one can prove that $\partial_2 (\gamma_1 + \gamma_4) = C_1^8 - C_1^7$. It is thus possible to show that the torus $T^2$ is associated with ${\cal H}_1 = \{ {C}_1^5, {C}_1^7 \}$, that is, it has two 1-generators. 

Since there are no 3-cells in a complex of dimension 2, ${\cal B}_2 (K(T^2)) = \emptyset$. 
Using that ${C}_1^5 \sim {C}_1^9-{C}_1^6$, one can show that $\partial_2 (\gamma_1 + \gamma_2 + \gamma_3 + \gamma_4) = 0$: this 2-chain is 
therefore a 2-cycle. It can be shown that all the 2-cycles are homologous to 
$\gamma_1 + \gamma_2 + \gamma_3 + \gamma_4$. Consequently, ${\cal H}_2 (K(T^2)) = \{ \displaystyle \gamma_1 + \gamma_2 + \gamma_3 + \gamma_4 \}$: it is made of a single 2-cycle. The Betti numbers for a torus are thus $\beta_0 
= 1$, $\beta_1 = 2$, and $\beta_2 = 1$, meaning that a torus T$^2$ is a single 
connected set, with two 1-generators, and one cavity.

To sum up:



\begin{eqnarray*}
{\cal H}_0 &=& [[ \langle1\rangle ]]  \\
{\cal H}_1 &=& [[ \langle 1,4\rangle - \langle 1,5\rangle 
	+ \langle 4,5 \rangle, 
	\langle 1,2 \rangle - \langle 1,3\rangle + \langle 2,3\rangle]] \\
{\cal H}_2 &=& [[ \gamma_1 + \gamma_2 + \gamma_3 +\gamma_4 ]]
\end{eqnarray*}

Since $K(T^2)$ is uniformly oriented, additional properties can be extracted. Here:  $$\partial({\Gamma})=\partial\left({\sum_i \gamma_i}\right)=0$$ since when the boundary operator is applied to the sum of all the 2-cells, the totality of the 1-cells cancel each other. This means that the torus has no boundaries: in other words, all the 1-cells in the contour of the complex in its planar representation are matched, without torsions. On the other hand, as $\partial{\Gamma}$ is null, there are no torsion elements.

\section{From the direct template to the reduced one}

The direct template drawn in Fig.\ \ref{bramaBS}(a) can be described by the 
linking matrices
\begin{equation}
  L_{ij} = 
  \left[
    \begin{array}{cc}
      +1 & +1 \\
      +1 & +1 
    \end{array}
  \right|
  \oplus
  \left|
    \begin{array}{cc}
      +1 & 0 \\
       0 & 0 
    \end{array}
  \right\rsem
  \otimes
  \left[
    \begin{array}{cc}
      +1 & +1 \\
      +1 & +1 
    \end{array}
  \right|
  \oplus
  \left|
    \begin{array}{cc}
      +1 &  0 \\
       0 &  0 
    \end{array}
  \right\rsem
\end{equation}
Combining each global torsion with the next mixer, we get
\begin{equation}
  L_{ij} = 
  \left[
    \begin{array}{cc}
      +1 &  +1 \\
      +1 & +2 
    \end{array}
  \right\rsem
  \otimes
  \left[
    \begin{array}{cc}
      +1 & +1 \\
      +1 & +2 
    \end{array}
  \right\rsem
\end{equation}
To transfom a multiplication into an addition between linking matrices, it is
necessary to extend each matrix into a $4 \times 4$ matrix with some rules as
follows.\cite{Ros15} The first matrix is simply expanded by replacing each of
its element $L_{ij}$ into a $2 \times 2$ submatrix whose elements are copies 
of the initial element $L_{ij}$. The second matrix is obtained by replacing 
each of its elements $L_{ij}$ with a block $B$ corresponding to the initial 
$2 \times 2$ second linking matrix transformed as\cite{Ros15}
\begin{equation}
  \label{expond}
	L_{ij}^{\rm E} =
  \left|
    \begin{array}{ll}
      B & L_{ii} \mbox{ and } L_{jj} \mbox{ are both even;}
        \\[0.1cm]
      B^\_ & L_{ii} \mbox{ is odd and } L_{jj} \mbox{ is even;}
        \\[0.1cm]
      B^| & L_{ii} \mbox{ is even and } L_{jj} \mbox{ is odd;}
        \\[0.1cm]
      B^p & L_{ii} \mbox{ and } L_{jj} \mbox{ are both odd.}
    \end{array}
  \right.
\end{equation}
where

\begin{itemize}

\item $B^\_\_$ is matrix $B$ whose row order is reversed;

\item $B^|$ is matrix $B$ whose column order is reversed;

\item $B^p$ is matrix $B$ which was permuted.

\end{itemize}
In the present case, the linking matrix $L^{\rm E}_{ij}$ is thus
expanded as
\begin{equation}
  L_{ij}^{\rm E} =
  \left[
    \begin{array}{cc}
       B^{\rm p} &  B^{\_\_} \\[0.1cm]
       B^{|} &  B 
    \end{array}
  \right\rsem
  =
  \left[
    \begin{array}{cccc}
      2 & 1 & 1 & 2 \\[0.1cm]
      1 & 1 & 1 & 1 \\[0.1cm]
      1 & 1 & 1 & 1 \\[0.1cm]
      2 & 1 & 1 & 2
    \end{array}
  \right\rsem
\end{equation}

\begin{equation}
  \begin{array}{rl}
	L_{ij}^{\rm E} &
	\displaystyle =
      \left[
        \begin{array}{cccc}
          1 & 1 & 1 & 1 \\[0.1cm]
          1 & 1 & 1 & 1 \\[0.1cm]
          1 & 1 & 2 & 2 \\[0.1cm]
          1 & 1 & 2 & 2
         \end{array}
       \right| +
      \left|
        \begin{array}{cccc}
           0 &  0 &  0 &  0 \\[0.1cm]
           0 &  0 &  0 &  1 \\[0.1cm]
           0 &  0 &  0 &  0 \\[0.1cm]
           0 &  1 &  0 &  0
         \end{array}
       \right| +
       \left[
         \begin{array}{cccc}
           2 & 1 & 1 & 2 \\[0.1cm]
           1 & 1 & 1 & 1 \\[0.1cm]
           1 & 1 & 1 & 1 \\[0.1cm]
           2 & 1 & 1 & 2
         \end{array}
       \right\rsem
	  \\[0.4cm]
	  & = \displaystyle 
      \left|
        \begin{array}{cccc}
           3 &  2 &  2 &  3 \\[0.1cm]
           2 &  2 &  2 &  3 \\[0.1cm]
           2 &  2 &  3 &  3 \\[0.1cm]
           3 &  3 &  3 &  4
         \end{array}
       \right\rsem
  \end{array}
\end{equation}
where the second matrix is a permutation matrix added to restore the correct 
branching between the different strips (see Rosalie and Letellier,\cite{Ros15}
for details). This linking matrix is in agreement with the matrix proposed for
this Burke and Shaw attractor.\cite{Let96a}



\bibliography{SysDyn}

\end{document}